\documentclass[preprint,journal]{vgtc}

\preprinttext{}

\makeatletter
\renewcommand{\teaser}[1]{%
  \renewcommand{\vgtc@teaser}{%
    \begin{minipage}{\linewidth}#1\end{minipage}%
    \setlength{\captionmargin}{0in}%
  }%
}
\patchcmd{\@maketitle}{0.33in}{0in}{}{\ClassError{arxiv}{Patching the abstract width failed}{}}
\makeatother
\manuscriptnote{}

\newcommand{\aff}[1]{\texorpdfstring{\textsuperscript{#1}}{}}

\PassOptionsToPackage{svgnames}{xcolor} 
\newcommand{\noMinted}{}                

\usepackage{amsmath,amsfonts}
\usepackage{array}
\usepackage{booktabs}
\usepackage{url}
\usepackage{verbatim}
\usepackage{graphicx}

\usepackage[T1]{fontenc}
\usepackage{listings}
\usepackage{xcolor}

\lstdefinelanguage{JavaScript}{
  basicstyle=\ttfamily\footnotesize,
  keywords={break, case, catch, continue, debugger, default, delete, do, else,
      finally, for, function, if, in, instanceof, new, return, switch, this, throw,
      try, typeof, var, void, while, with, let, const, yield, async, await, of,
      class, extends, super, import, export, null, true, false, undefined},
  morecomment=[l]{//},
  morecomment=[s]{/*}{*/},
  morestring=[b]',
  morestring=[b]",
  sensitive=true,
  showstringspaces=false,
  breaklines=true,
  ndkeywords={await, async, const, let, var, function, class, extends, return,
      this, new, if, else, for, while, do, switch, case, break, continue, try,
      catch, finally, throw, typeof, instanceof, in, of, import, export, from,
      default, delete, void, yield, super},
  keywordstyle=\color{blue}\bfseries,
  ndkeywordstyle=\color{darkblue}\bfseries,
  identifierstyle=\color{black},
  commentstyle=\color{green!60!black}\itshape,
  stringstyle=\color{red!70!black},
}

\definecolor{punct}{RGB}{0,0,180}
\definecolor{delim}{RGB}{20,105,176}
\definecolor{keycolor}{RGB}{180,0,0} 

\lstdefinelanguage{json}{
basicstyle=\ttfamily\footnotesize,
showstringspaces=false,
breaklines=true,
backgroundcolor=\color{gray!10},
frame=none,
numbers=none,
literate=
{\"}{{{\color{keycolor}{\"}}}}{1}
{:}{{{\color{punct}{:}}}}{1}
{,}{{{\color{punct}{,}}}}{1}
{\{}{{{\color{delim}{\{}}}}{1}
{\}}{{{\color{delim}{\}}}}}{1}
{[}{{{\color{delim}{[}}}}{1}
{]}{{{\color{delim}{]}}}}{1}
{0}{{{\color{blue}0}}}{1}
{1}{{{\color{blue}1}}}{1}
{2}{{{\color{blue}2}}}{1}
{3}{{{\color{blue}3}}}{1}
{4}{{{\color{blue}4}}}{1}
{5}{{{\color{blue}5}}}{1}
{6}{{{\color{blue}6}}}{1}
{7}{{{\color{blue}7}}}{1}
{8}{{{\color{blue}8}}}{1}
{9}{{{\color{blue}9}}}{1}
}

\ifdefined\noMinted\else
\usepackage{minted}
\immediate\write18{bash ./pygmentize_patcher.sh}
\fi

\usepackage{url}
\usepackage{enumitem}                  

\usepackage{pgfplots}
\pgfplotsset{compat=1.17}
\usetikzlibrary{arrows.meta}

\definecolor{likertOne}{HTML}{B93B3A}
\definecolor{likertTwo}{HTML}{EE8E8A}
\definecolor{likertThree}{HTML}{E2E1DA}
\definecolor{likertFour}{HTML}{86B6EF}
\definecolor{likertFive}{HTML}{256ABF}
\definecolor{chartInk}{HTML}{0B0B0B}
\definecolor{chartSecondary}{HTML}{52514E}
\definecolor{chartMuted}{HTML}{898781}
\definecolor{chartGrid}{HTML}{E1E0D9}
\definecolor{chartBaseline}{HTML}{C3C2B7}

\pgfplotsset{
  likert bar chart/.style={
      xbar stacked,
      width=\linewidth,
      xmin=0, xmax=10,
      xtick={0,2,...,10},
      xlabel={Number of participants},
      xlabel style={font=\scriptsize, color=chartSecondary},
      ytick=data,
      yticklabel style={font=\footnotesize, color=chartInk},
      xticklabel style={font=\scriptsize, color=chartMuted},
      tick style={draw=none},
      axis x line*=bottom,
      axis y line*=left,
      y axis line style={draw=none},
      x axis line style={chartBaseline, line width=0.5pt},
      xmajorgrids,
      grid style={chartGrid, line width=0.4pt},
      bar width=9pt,
      legend style={
          at={(0.5,1.02)}, anchor=south,
          legend columns=-1, draw=none,
          font=\scriptsize,
          /tikz/every even column/.append style={column sep=0.5em},
        },
      nodes near coords={\pgfmathfloatifflags{\pgfplotspointmeta}{0}{}{\pgfmathprintnumber{\pgfplotspointmeta}}},
      every node near coord/.append style={font=\scriptsize},
    },
  likert level/.style 2 args={fill=#1, draw=white, line width=0.8pt,
      every node near coord/.append style={text=#2}},
  likert1/.style={likert level={likertOne}{white}},
  likert2/.style={likert level={likertTwo}{chartInk}},
  likert3/.style={likert level={likertThree}{chartInk}},
  likert4/.style={likert level={likertFour}{chartInk}},
  likert5/.style={likert level={likertFive}{white}},
}

\usepackage[normalem]{ulem}

\usepackage[most]{tcolorbox}
\newtcolorbox{todobox}{
  colback=red!20,
  colframe=red!80!black,
  title=TODO,
  sharp corners,
  boxrule=0.8pt
}

\makeatletter
\if@twocolumn
  \newcommand{\smallfigwidth}{\columnwidth}
\else
  \newcommand{\smallfigwidth}{0.7\columnwidth}
\fi
\makeatother

\title{Vibe Coding Reality: Rapid Prototyping of Spatially Aware Augmented Reality Applications}

\author{%
  \authororcid{Jan Kolberg\aff{1}}{0009-0006-8106-4429},
  \authororcid{Michael Pabst\aff{1}}{0000-0002-7985-4652},
  \authororcid{Shohei Mori\aff{1}}{0000-0003-0540-7312},
  and \authororcid{Dieter Schmalstieg\aff{1,2}}{0000-0003-2813-2235}
}

\affiliation{%
  \normalsize
  \textsuperscript{1}VISUS, University of Stuttgart, Germany \qquad
  \textsuperscript{2}Graz University of Technology, Austria
}

\authorfooter{
  \item E-mail: \{first.last\}@visus.uni-stuttgart.de.
  \item J. Kolberg is the corresponding author.
}

\abstract{%
  Vibe coding, the creation of software through natural-language prompts, has transformed application prototyping. However, Augmented Reality (AR) has remained out of reach: Coding agents know nothing about the physical environment an AR application will run in. We close this gap by combining a semantic scene graph with a vibe coding agent. While writing code, the agent queries this structure for objects, spatial and semantic relationships. The vibe-coded application inspects the live environment model at runtime to stay consistent with reality. To demonstrate versatility, we recreate 16 published AR applications and evaluate 120 vibe coding examples. We report an expert study in which 10 participants created 40 spatially registered applications using devices connected to the Internet of Things. Overall, 87.5\% of the participants' applications were rated successful at a median of two prompts, with a mean SUS score of 85, confirming the system as a rapid prototyping tool for situated AR.
}


\teaser{
  \centering
  \includegraphics[width=\linewidth]{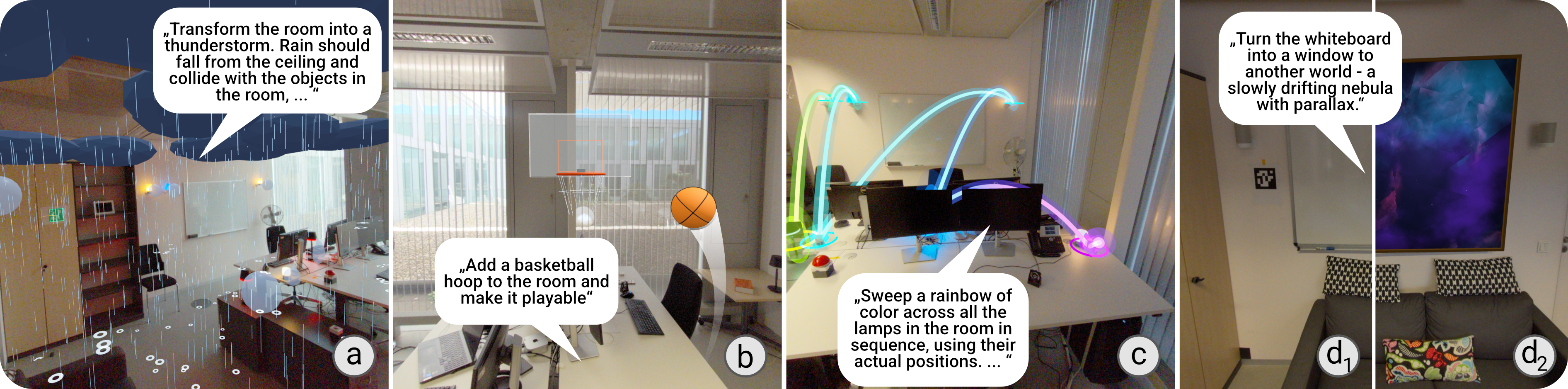}
  \caption{We asked our AR vibe coding agent to create (a) thunderstorm with IoT lamp activation during lightning strikes, colliding water droplets with the environment and generated sounds, (b) playable basketball with scoring logic, where a hand-tracked virtual ball bounces off the virtual hoop net and the surrounding real environment (a trail was added to highlight the motion of the ball), (c) sequentially animated IoT lamps through virtual rainbow animation, and (d) animated wall portal, shown without ($\text{d}_1$) and with ($\text{d}_2$) the application.}
  \label{fig:teaser}
}

\begin{document}

\firstsection{Introduction}

\maketitle

\textit{Vibe coding} is the process of creating functional software from text prompts using a large language model (LLM) without writing code~\cite{fawzy2025vibecoding, lovable2026platform, bolt2026stackblitz, v02026vercel}. Today, entire applications, from mobile apps to full websites, can be produced by users with no programming background by describing the desired behavior to an LLM~\cite{google2026antigravity, v02026vercel}. This capability has dramatically lowered the barrier for non-experts to developing their own software.

Despite its popularity, when it comes to extended reality (XR) and especially augmented reality (AR), vibe coding is in its infancy. Creating AR content has long been recognized as an activity with a steep learning curve, even for experienced developers~\cite{ashtari2020creating, nebeling2018trouble}. The vast majority of AR experiences are still built with heavyweight game engines such as Unity, Unreal, or Godot, which couple coding with manual workflows in a graphical editor: Scene hierarchies, component configurations, asset placement, and spatial relationships must be authored by hand and can hardly be produced by an LLM in the same manner as, for example, a C\# script. This tight coupling between coding and manual scene authoring fundamentally limits the extent to which AR applications can be vibe coded.

Recent advances in WebXR~\cite{w3c2026webxr} make it possible to create XR experiences that run directly in the browser using pure JavaScript, without the need for a dedicated game engine~\cite{Li2025XR}. The purely textual representation of a WebXR application is better suited for generation with an LLM than the binary assets required by game engines.

However, for AR, an essential ingredient is still missing. AR applications must be situated in and aware of the user's physical surroundings~\cite{schmalstieg2016augmented}. Unfortunately, current vibe-coding agents have no model of the environment in which the application will run. In the best case, they see a single camera image~\cite{zhu2025agentar} or the coarse planes that WebXR detects~\cite{du2026vibe} for planning. The agents do not know which physical objects are present in the room, where they are located, or which of them are controlled via the Internet of Things (IoT). As a result, the vibe-coded applications are spatially blind: Virtual content cannot be spatially registered in the environment. More advanced context-aware AR, such as placing virtual controls next to an electronic device or reasoning about the feasibility of a particular user request given the current room, are out of reach.

In this work, we close this gap by combining a \textit{semantic scene graph} (SSG) with a vibe coding agent that targets the WebXR ecosystem~\cite{w3c2026webxr}. An SSG is a structured representation of the physical environment that represents objects, their geometry, as well as their spatial and semantic relationships~\cite{armeni20193d, rosinol2021kimera, gu2024conceptgraphs}.
Figure~\ref{fig:teaser} shows examples of applications generated with our system. In summary, our contributions are:

\begin{itemize}[leftmargin=*, itemsep=2pt, parsep=0pt, topsep=2pt]
  \item We supply an SSG to a vibe coding framework to empower it with reasoning abilities about the spatial environment. Our system integrates tools that allow querying the SSG both at generation time and at runtime for context-sensitive application behavior.
  \item We introduce a validation mechanism that does not only statically check code correctness, but also plausibility in the physical environment (such as checks on the existence of required objects).
  \item We perform a technical evaluation of the vibe-coded AR applications using a mixture of 120 existing and new spatially focused test cases.
  \item We replicate 16 AR applications from the literature to demonstrate the versatility of our approach and its ability to match manual authoring.
  \item We report on a qualitative interview study involving 10 domain experts, in which the effectiveness of our approach was evaluated from a practical perspective.
\end{itemize}

\section{Related work}

We discuss the state of the art in the two areas that are fundamental to our work: semantic scene graphs, and LLM agents for XR applications.

\subsection{Semantic scene graphs}

A scene graph is an established representation in computer graphics that describes a scene through its objects, their attributes, and their hierarchical relationships~\cite{Strauss1992}. When machine learning recently acquired the ability to extract semantic relationships between objects in an image~\cite{johnson2015image}, the term ``scene graph'' was adopted by the computer vision community and given a broader meaning. Armeni et al.~\cite{armeni20193d} lifted the scene graph to 3D by organizing buildings, rooms, objects, and cameras in a hierarchical graph. Subsequent work established the term ``3D scene graph'' as a sparse, semantic representation of physical environments for intelligent agents~\cite{kim2020scenegraph}; to avoid any ambiguities, we refer to this idea as a \textit{semantic scene graph} or \textit{SSG} in this paper.

Wald et al.~\cite{wald2020learning} created the 3DSSG dataset and a learned model that regresses an SSG, including support and proximity relationships, directly from point clouds. Recent approaches construct the SSG online: SceneGraphFusion~\cite{wu2021scenegraphfusion} incrementally predicts the graph from RGB-D streams, while Kimera~\cite{rosinol2021kimera} and Hydra~\cite{hughes2022hydra} build hierarchical scene graphs from metric-semantic SLAM in real time. However, these approaches use a closed set of labels.

More recently, vision-language foundation models removed this restriction. ConceptGraphs~\cite{gu2024conceptgraphs}, Open3DSG~\cite{koch2024open3dsg}, and ThinkGraphs~\cite{bickici2026think} produce open-vocabulary scene graphs that can be queried in natural language. SceneScript~\cite{avetisyan2024scenescript} expresses the scene structure as an autoregressive language of commands, and functional scene graphs~\cite{zhang2025openvocab} even capture functional elements such as switches and appliances. In XR, semantic scene graphs have been used to adapt predefined content to the physical world: Retargetable AR~\cite{tahara2020retargetable} matches an abstract target scene graph against the graph of the user's room. Storytelling guided by scene semantics places narrative events on suitable real objects~\cite{li2022storytelling}. LORIOT~\cite{kolberg2026loriot} binds IoT devices to scene graph nodes for localization. In all these systems, the scene graph is consumed by a fixed pipeline or a predefined application; it is never exposed to a code-generating agent. We close this gap and make the SSG available to an LLM coding agent that targets  WebXR applications.

\subsection{LLM agents for XR applications}

A growing body of work turns natural language into content for XR. Prior research focuses on VR, because knowing the virtual environment in detail makes it easier to insert matching content. De~la~Torre et al.~\cite{de2024llmr} prompt interactive worlds in real time with LLMR. Chen et al.~\cite{chen2025llmer} construct interactive XR scenes from LLM-generated JSON. Yin et al.~\cite{yin2024text2vrscene} assemble entire VR experiences from a single text prompt. For static 3D environments, Holodeck~\cite{yang2024holodeck} pairs an LLM with a large 3D-asset library and layout optimization to populate rooms from a description. VRCopilot~\cite{zhang2024vrcopilot} keeps the user in a mixed-initiative loop for authoring 3D layouts in VR. DreamCodeVR~\cite{giunchi2024dreamcodevr} lets non-programmers change the behavior of a running VR application by translating speech into small code edits. GenAssist~\cite{srinidhi2026genassist} generates XR programs from prompts with the help of retrieval augmented generation to access collections of assets and scripts.

More recent work targets AR and combines generative AI with knowledge of the real world. AgentAR~\cite{zhu2025agentar} uses an LLM that creates AR applications in Unity using natural language and a library of AR-specific tools. ImaginateAR~\cite{lee2025imaginatear} combines offline scene understanding with rapid 3D-asset generation and speech interaction for situated AR authoring. CARING-AI~\cite{shi2025caringai} uses generative AR to create context-aware instructions using avatars in AR. CinemaWorld~\cite{ihara2026cinemaworld} supplements movie consumption with ambient visuals, which are generated by an LLM, synchronized with the movie track and displayed in AR. XR-Blocks~\cite{Li2025XR, du2026vibe} shows that WebXR development can leverage vibe coding. However, all of these works generate content without (or negligible) knowledge of the physical environment. Some systems make use of the depth sensor or plane detection to place virtual objects on the ground, but none takes physical objects into consideration. Our approach departs from this pattern by constructing a vibe-coding module that explicitly grounds program generation in environmental data, incorporating details about smart devices alongside other spatial features.

A final category of research relies on an LLM to interpret spatial contexts on the fly, delivering contextual data immediately before AR interactions occur. XaiR~\cite{srinidhi2024xair} bridges an LLM with a tangible environment, allowing users to communicate with physical items via natural-language commands. Satori~\cite{li2025satori} provide context-aware guidance by integrating a multimodal LLM with a user model based on a belief--desire--intention structure. SemanticReality~\cite{liu2026} uses a set of LLM prompts to inquire about the specific relationships of objects detected in the current environment. These relationships resemble a subset of edges in an SSG, but queries in real time are limited to a small window of recent objects. In contrast to this previous work, we acquire a persistent representation of the physical environment in the form of an SSG. It aggregates both geometric and semantic information and allows for full context awareness in content generation via LLM tools.

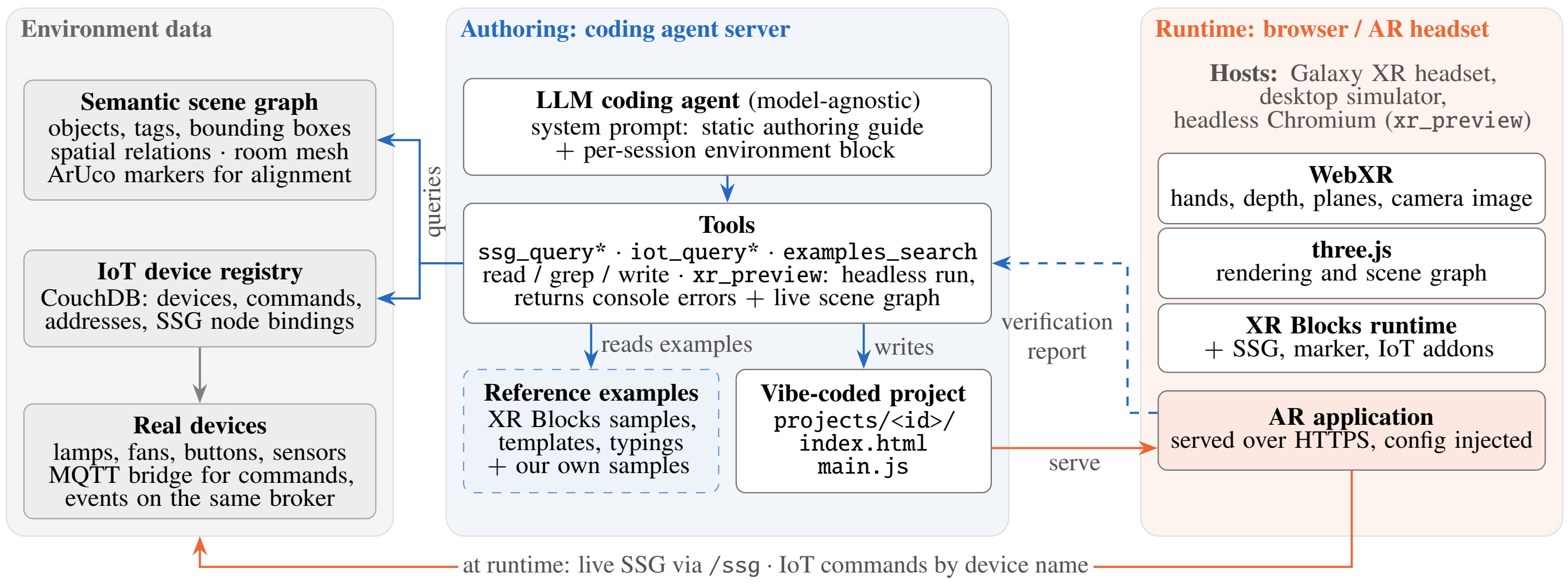
\begin{figure*}[t]
  \centering
  \resizebox{\linewidth}{!}{
%
%
\definecolor{flowRuntime}{HTML}{EB6834}
\begin{tikzpicture}[
    font=\footnotesize,
    box/.style={rounded corners=3pt, align=center, inner sep=4pt, draw=black!55, fill=white},
    envbox/.style={box, fill=black!7, draw=black!35, text width=3.72cm},
    authbox/.style={box, text width=5.72cm},
    halfbox/.style={box, text width=2.62cm, minimum height=1.4cm},
    repo/.style={fill=likertFive!8, draw=likertFive!70, dashed},
    rtbox/.style={box, text width=4.12cm},
    hull/.style={rounded corners=6pt, draw=black!12, line width=0.4pt},
    hulllabel/.style={font=\scriptsize\bfseries, anchor=north west},
    flow/.style={-{Stealth[length=2.2mm]}, line width=0.7pt},
    sidelabel/.style={font=\scriptsize, text=chartSecondary, inner sep=1pt, align=center},
    inlinelabel/.style={font=\scriptsize, text=chartSecondary, inner sep=1.5pt, fill=white},
  ]

  \draw[hull, fill=black!4]       (0.0, 0.3) rectangle (4.4, -5.7);
  \draw[hull, fill=likertFive!6]  (5.0, 0.3) rectangle (11.4, -5.7);
  \draw[hull, fill=flowRuntime!8] (12.9, 0.3) rectangle (17.7, -5.7);
  \node[hulllabel, text=black!60]    at (0.05, 0.28)  {Environment data};
  \node[hulllabel, text=likertFive]  at (5.05, 0.28)  {Authoring: coding agent server};
  \node[hulllabel, text=flowRuntime] at (12.95, 0.28) {Runtime: browser / AR headset};

  \node[envbox] (ssg) at (2.2, -1.2) {\textbf{Semantic scene graph}\\[-1pt]
    \scriptsize objects, tags, bounding boxes\\[-2pt]
    \scriptsize spatial relations \textperiodcentered{} room mesh\\[-2pt]
    \scriptsize ArUco markers for alignment};
  \node[envbox] (iot) at (2.2, -3.0) {\textbf{IoT device registry}\\[-1pt]
    \scriptsize CouchDB: devices, commands,\\[-2pt]
    \scriptsize addresses, SSG node bindings};
  \node[envbox] (devices) at (2.2, -4.85) {\textbf{Real devices}\\[-1pt]
    \scriptsize lamps, fans, buttons, sensors\\[-2pt]
    \scriptsize MQTT bridge for commands,\\[-2pt]
    \scriptsize events on the same broker};
  \draw[flow, black!50] (iot) -- (devices);

  \node[authbox] (agent) at (8.2, -1.05) {\textbf{LLM coding agent} \scriptsize (model-agnostic)\\[-1pt]
    \scriptsize system prompt: static authoring guide\\[-2pt]
    \scriptsize $+$ per-session environment block};
  \node[authbox] (tools) at (8.2, -2.6) {\textbf{Tools}\\[-1pt]
    \scriptsize \texttt{ssg\_query*} \textperiodcentered{} \texttt{iot\_query*} \textperiodcentered{} \texttt{examples\_search}\\[-2pt]
    \scriptsize read / grep / write \textperiodcentered{} \texttt{xr\_preview}: headless run,\\[-2pt]
    \scriptsize returns console errors $+$ live scene graph};
  \node[halfbox, repo, anchor=north] (refex) at (6.65, -3.8) {\textbf{Reference examples}\\[-1pt]
    \scriptsize XR Blocks samples,\\[-2pt]
    \scriptsize templates, typings\\[-2pt]
    \scriptsize $+$ our own samples};
  \node[halfbox, anchor=north] (project) at (9.75, -3.8) {\textbf{Vibe-coded project}\\[-1pt]
    \scriptsize \texttt{projects/<id>/}\\[-2pt]
    \scriptsize \texttt{index.html}\\[-2pt]
    \scriptsize \texttt{main.js}};

  \draw[flow, likertFive] (agent) -- (tools);
  \draw[flow, likertFive] (refex.north |- tools.south) -- (refex.north)
    node[sidelabel, midway, right=2pt] {reads examples};
  \draw[flow, likertFive] (project.north |- tools.south) -- (project.north)
    node[sidelabel, midway, right=2pt] {writes};

  \draw[flow, likertFive] (tools.west) -- (4.7, -2.6) -- (4.7, -1.2) -- (ssg.east);
  \draw[flow, likertFive] (4.7, -2.6) -- (4.7, -3.0) -- (iot.east);
  \node[sidelabel, rotate=90] at (4.87, -1.9) {queries};

  \node[font=\scriptsize, align=center, text=chartSecondary, text width=4.12cm] (hosts) at (15.3, -0.72)
    {\textbf{Hosts:} Galaxy XR headset,\\[-2pt]
     desktop simulator,\\[-2pt]
     headless Chromium (\texttt{xr\_preview})};
  \node[rtbox] (webxr) at (15.3, -1.75) {\textbf{WebXR}\\[-2pt]
    \scriptsize hands, depth, planes, camera image};
  \node[rtbox] (three) at (15.3, -2.6) {\textbf{three.js}\\[-2pt]
    \scriptsize rendering and scene graph};
  \node[rtbox] (xrb) at (15.3, -3.45) {\textbf{XR Blocks runtime}\\[-2pt]
    \scriptsize $+$ SSG, marker, IoT addons};
  \node[rtbox, fill=flowRuntime!15, minimum height=0.9cm] (app) at (15.3, -4.5) {\textbf{AR application}\\[-2pt]
    \scriptsize served over HTTPS, config injected};

  \draw[flow, flowRuntime] (project.east |- 0,-4.7) -- (app.west |- 0,-4.7);
  \node[sidelabel, anchor=north] at (12.15, -4.78) {serve};

  \draw[flow, likertFive, dashed] (app.west |- 0,-4.3) -- (12.75, -4.3) -- (12.75, -2.6) -- (tools.east);
  \node[sidelabel] at (11.95, -3.45) {verification\\report};

  \draw[flow, flowRuntime] (app.south) -- (15.3, -6.05) -- (2.2, -6.05)
    node[inlinelabel, midway] {at runtime: live SSG via \texttt{/ssg} \textperiodcentered{} IoT commands by device name}
    -- (2.2, -5.7);
\end{tikzpicture}}
  \caption{System architecture. \textbf{Environment data} (gray): The semantic scene graph provides a representation of physical objects, relational connections, and the room mesh. It also includes an IoT registry synchronized with corresponding SSG nodes, alongside actual hardware communicating through an MQTT gateway. \textbf{Authoring} (blue): The model-agnostic coding agent queries the surroundings using specialized tools. By examining sample projects, it extracts reliable implementation strategies before generating the complete codebase. \textbf{Runtime} (orange): Deployment relies on a combined WebXR, three.js, and XR Blocks foundation across XR headsets, desktop emulation environments, or in a headless-browser evaluation (\texttt{xr\_preview}). The application launches with dynamically supplied parameters. Besides static employment, the SSG and IoT data can further be queried during execution, while the headless evaluation relays console logs back to the agent.}
  \label{fig:data_access}
\end{figure*}

\begin{figure}[t]
  \centering
  \includegraphics[width=\smallfigwidth,height=0.72\textheight,keepaspectratio]{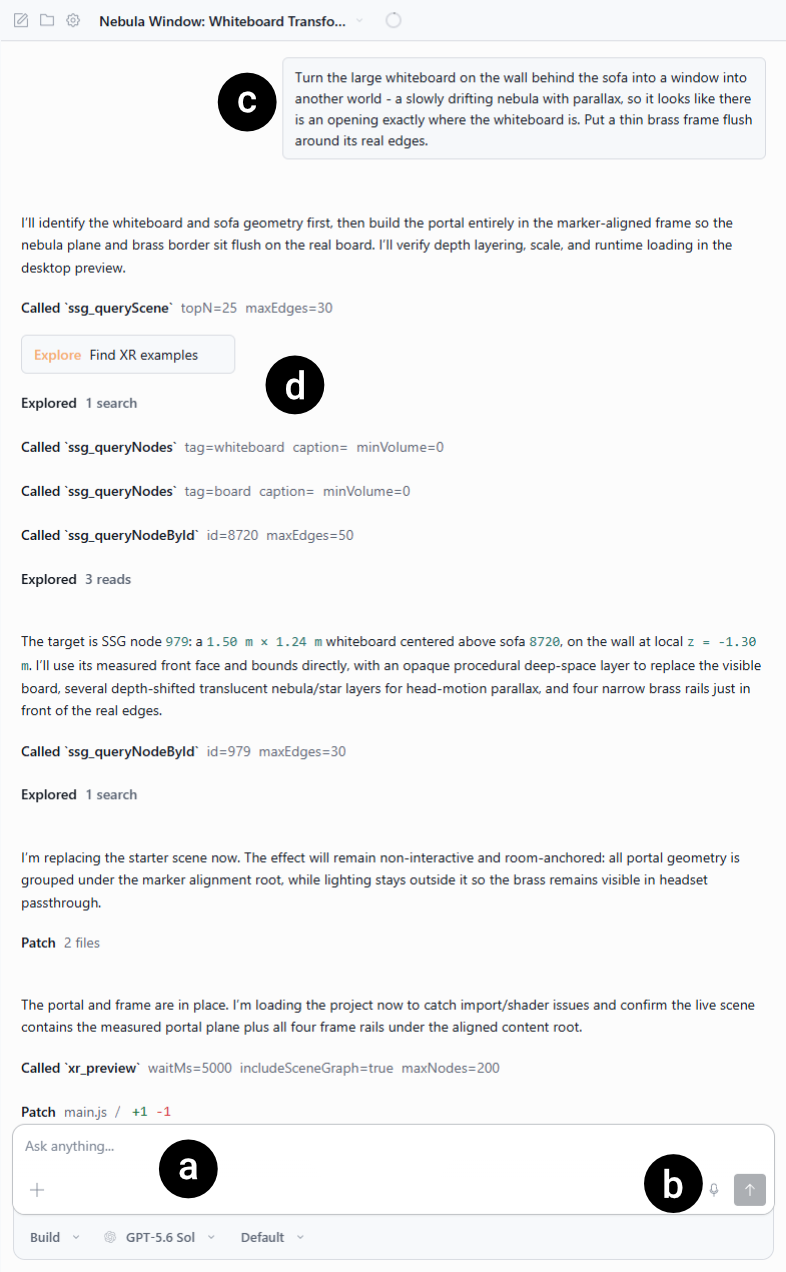}
  \caption{Chat interface of the coding agent, which can be opened in a browser either directly on an AR headset or on a desktop computer. Input interface with either typed (a) or voice (b) requests, (c) agent request, (d) agent reasoning response, including several tool calls and their individual outputs.}
  \label{fig:coding_agent_interface}
\end{figure}

\section{Approach}
\label{sec:implementation}

The specific challenge of vibe-coding for XR applications comes from requirements for tools and resources that differ radically from traditional domains (e.g., data processing). XR applications must map low-level sensor events, such as hand detection, to high-level activities. The logic of XR applications is often strongly dependent on the presence, variety and layout of scene objects. It is common that scene objects need to be generated or arranged to match the goals of the application. This ``content generation'' step of 3D applications is traditionally carried out with an interactive editor, but, for vibe coding, needs to be performed automatically.

Conventional workflows for XR development, which rely on binary formats for visual content and use languages that are compiled ahead of time, such as C\#, are not particularly well suited for vibe coding~\cite{chen2025llmer}. The WebXR~\cite{w3c2026webxr} ecosystem offers an alternative based on a standardized API for XR applications running in the web-browser. WebXR addresses several of the mentioned challenges. It offers a convenient abstraction of the low-level details of XR devices. It relies on textural representations, such as markup languages and JavaScript code, which is compiled at runtime and relatively forgiving to imperfect code due to its weak type system. Application developers rely on JavaScript libraries such as three.js or babylon.js, which combine WebXR with WebGPU for graphics output, and expose a high-level interface in the form of a hierarchical scene graph (the traditional computer graphics variety) to the application developer.

XR-Blocks~\cite{Li2025XR} fills the remaining gaps for successful vibe coding of XR applications with  high-level abstractions. It provides JavaScript components on top of three.js, which represent the user's body and the physical environment, and are easily integrated with LLM tools. However, XR-Blocks relies on the sensing abilities of WebXR, which are restricted to hand input and raw depth sensing. Beyond this limited sensing, applications have no notion of the physical world. Consequently, AR applications can only present virtual content that is spatially anchored to physical coordinates, but has no access to any semantic information associated with these coordinates. In the terminology of Willet et al.~\cite{willett_embedded_2017}, the applications are \textit{immersive, but not situated}.

Our vibe-coding solution is unique in providing a mapping from the virtual environment of the XR application to the physical environment (Figure~\ref{fig:data_access}). For that purpose, we provide an SSG which describes the objects in the environment, their attributes and their relationships. An SSG was originally designed to let robots reason about their physical environment; we repurpose it to perform a similar reasoning for vibe coding. By querying the SSG, a vibe-coded application can perform feats such as highlighting the objects in the room that are relevant to a task or placing a virtual control panel right next to the physical device that it operates. For structuring the virtual scene and spatial interaction, we adopt the XR-Blocks library~\cite{Li2025XR}. However, we use our own vibe-coding solution (the vibe-coding solution described by Du et al.~\cite{du2026vibe} is not publicly available and would not be compatible with out SSG).

The remainder of this section explains the various technical components of our approach:
Section~\ref{sec:system_overview} gives an overview of the system architecture and its main components.
Section~\ref{sec:agent} introduces the vibe-coding agent.
Section~\ref{sec:agent_tooling} lists the tools through which the agent analyzes the physical environment. Section~\ref{sec:verification} explains the verification process used by the agent to produce robust application code.
Finally, Section~\ref{sec:spatial_grounding} discusses the spatial object grounding mechanism.

\begin{figure}[t]
  \centering
%
\definecolor{flowRuntime}{HTML}{EB6834}
\begin{tikzpicture}[
    font=\footnotesize,
    box/.style={rounded corners=3pt, align=center, inner sep=4pt, minimum width=7.4cm},
    source/.style={box, fill=black!7, draw=black!35, minimum height=0.95cm},
    stage/.style={box, fill=white, draw=black!55, minimum height=0.95cm},
    deliver/.style={box, fill=likertFour!30, draw=black!45, minimum height=0.8cm},
    flow/.style={-{Stealth[length=2.2mm]}, line width=0.7pt},
    flowlabel/.style={font=\scriptsize, text=chartSecondary},
  ]

  \fill[likertFive!6, rounded corners=5pt] (-4.1, -1.125) rectangle (4.1, -4.975);
  \node[font=\scriptsize, text=likertFive, anchor=south east]
  at (4.1, -1.09) {one tool-augmented loop};

  \node[source] (prompt) at (0, 0) {\textbf{System prompt}\\[-1pt]
    \scriptsize static authoring guide $+$ per-session \texttt{<env>}};
  \node[stage] (ground) at (0, -1.45) {\textbf{Ground}\\[-1pt]
    \scriptsize scene overview \textperiodcentered{} nodes/edges \textperiodcentered{} IoT registry \textperiodcentered{} examples};
  \node[stage] (write) at (0, -2.90) {\textbf{Write}\\[-1pt]
    \scriptsize \texttt{index.html} $+$ \texttt{main.js} into the session folder};
  \node[stage] (verify) at (0, -4.35) {\textbf{Verify}\\[-1pt]
    \scriptsize console errors \textperiodcentered{} live scene graph \textperiodcentered{} request checklist};
  \node[deliver] (deliver) at (0, -5.9) {\textbf{Serve to the headset}\\[-1pt]
    \scriptsize with injected runtime configuration};

  \draw[flow] (0, 1.0) -- (prompt.north)
  node[flowlabel, midway, right] {user request};

  \draw[flow, likertFive] (prompt) -- (ground);
  \draw[flow, likertFive] (ground) -- (write);
  \draw[flow, likertFive] (write) -- (verify);
  \draw[flow, flowRuntime] (verify) -- (deliver)
  node[flowlabel, midway, right, align=left] {clean \&\\[-2pt]complete};

  \draw[flow, likertFive] (verify.west) -- ++(-0.3, 0) |- (write.west);
  \node[flowlabel, anchor=west] at (-3.75, -3.625) {edit \& re-verify};
\end{tikzpicture}
  \caption{The agent grounds the request in the room model, writes the project, and validates it by running it. Failure to validate results in another cycle of refinement before the validated project is finally served to the headset. Follow-up requests enter the same loop with the already established physical configuration added as contextual information (see Section~\ref{sec:spatial_grounding} for a detailed explanation).}
  \label{fig:agent_loop}
\end{figure}
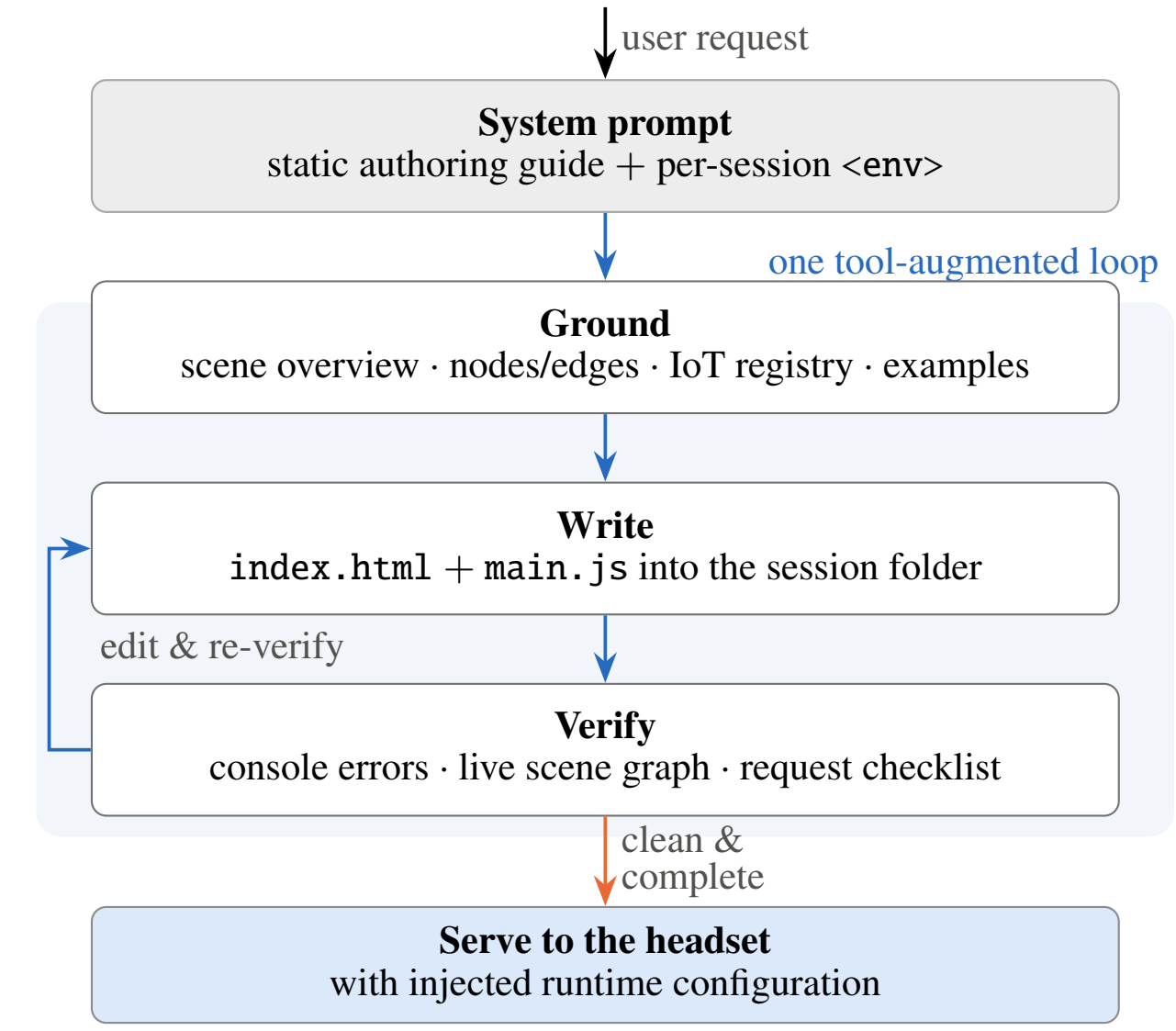

\subsection{System overview}
\label{sec:system_overview}

Our system is implemented using a client-server architecture. The server hosts the authoring interface (Figure~\ref{fig:coding_agent_interface}), runs the agent, and deploys the vibe-coded application to the client device. The client runs a WebXR-enabled browser, either on the headset (we use a Samsung Galaxy XR) or, during development, on a desktop machine with simulated headset inputs. Therefore, bootstrapping means loading a webpage which includes the required components (our SSG queries, three.js, XR-Blocks).

The SSG is stored in a database service accessible via HTTP. Therefore, the same SSG can be queried by the vibe-coding agent and later, at runtime, by the vibe-coded application, ensuring that the data is semantically consistent and up to date. Queries can retrieve scene objects, their attributes and their relationships. For IoT devices, the attributes stored in the SSG include not only their spatial position in the scene, but also the details of their digital interface, i.e., the format and semantics of the messages used to communicate with the IoT device. Our collection contains 26 IoT devices from various vendors as well as custom devices.

We spatially register the SSG with the coordinate system used by the ego-tracking service of the XR device  at runtime (Figure~\ref{fig:ssg_alignment}) using ArUco markers placed at known positions in SSG coordinates. The runtime library bundled with the vibe-coded client applications performs marker tracking and automatically aligns the coordinate systems of SSG and device tracking.

\subsection{Vibe-coding agent}
\label{sec:agent}

From a general point of view, an LLM transforms text into text. Compared to natural language text, program code is much more constrained, making it easy for an LLM to synthesize useful output~\cite{fawzy2025vibecoding}. However, a standalone LLM is usually not sufficient to solve specialized coding tasks. To achieve satisfactory results, the LLM must be embedded in an \textit{agentic framework} consisting of one or more agents.

An \textit{agent} uses the LLM as its reasoning engine, but does not simply return a text string. Instead, it solves a problem by iterating an analysis--action--evaluation loop. During analysis, it breaks complex goals into smaller tasks and formulates a plan. Then it takes actions by choosing and applying external tools. Finally, it evaluates the results and iterates with a refined plan if it determines that the problem has not yet been sufficiently solved.

In order to carry out this process, an agent needs to be provided with reasoning capability, memory, tools and resources. \textit{Reasoning} combines the user's prompt with techniques such as chain of thought~\cite{wei2022chainofthought}. The agent's \textit{memory} stores the plan and the history of activities. \textit{Tools} enable the agent to gather information about the environment or manipulate it. Finally, \textit{resources} are a collective term for instructions, API documentation and code examples, which are selectively included in the LLM prompt to enable in-context learning.

Our vibe-coding agent follows this pattern, as illustrated in Figure~\ref{fig:agent_loop}. Each user request invokes the LLM analysis with the prompt described in the following, supplemented with the accumulated history and schemas that describe all available tools. If the LLM requests actions involving tool calls, the server executes these calls, stores their results in its memory, evaluates the results, and invokes the LLM again. This process ends when the LLM responds without requesting further actions.

The prompt combines the user's request with a system prompt, which is itself composed of a small dynamic part (pointers to project resources, hardware configuration, physical coordinate system, date and time, etc.) and a larger static part that contains instructions for the agent. The static part instructs the agent to employ the following strategies:
\begin{itemize}
  \item The agent should inspect the physical environment with its tools if the goal requires it.
  \item The agent should generate code that queries the SSG service at runtime to obtain information such as object semantics, their spatial layout, or interface specification for IoT devices.
  \item The agent should use the XR-Blocks library for 3D interaction and various user interface components.
  \item The agent should look for suitable examples in the sample collection, but be aware that the examples may not adhere to certain mandatory coding standards.
  \item The agent should keep track of all sub-goals and elements (such as scene objects) present in the solution. It should run the validation tool (Section~\ref{sec:verification}) and iterate its solution until the validation is fully successful.
\end{itemize}
With these instructions, a typical vibe-coding process could progress as follows:
\textit{The user requests to ``place a virtual button on the nightstand, which can be pressed to toggle the ceiling lamp on or off''. The agent first retrieves the most relevant code sample (one demonstrating IoT control) and queries the scene overview to confirm that a nightstand and a lamp are present. It resolves the lamp reference through the IoT device tool to obtain its command interface and node id in the SSG. Using the node id, it inspects the nightstand's record to confirm that the placement is feasible, and generates code that places a virtual button on the nightstand's bounding box, while incorporating the IoT command interaction. }

\subsection{Tools and resources}
\label{sec:agent_tooling}

In addition to the standard tools for agentic software engineering~\cite{yao2023react, yang2024sweagent}---such as file manipulation, history collection, or summarization---we provide the agent with tools that enable it to interpret the layout of the physical environment and the capabilities of the IoT devices. The agent can look up these elements on demand and condition the generated code on the returned data. The tools expose the data sources above, together with a catalog of reference example projects, in a structured, queryable form:

\begin{figure}[t]
  \centering
  \includegraphics[width=\smallfigwidth]{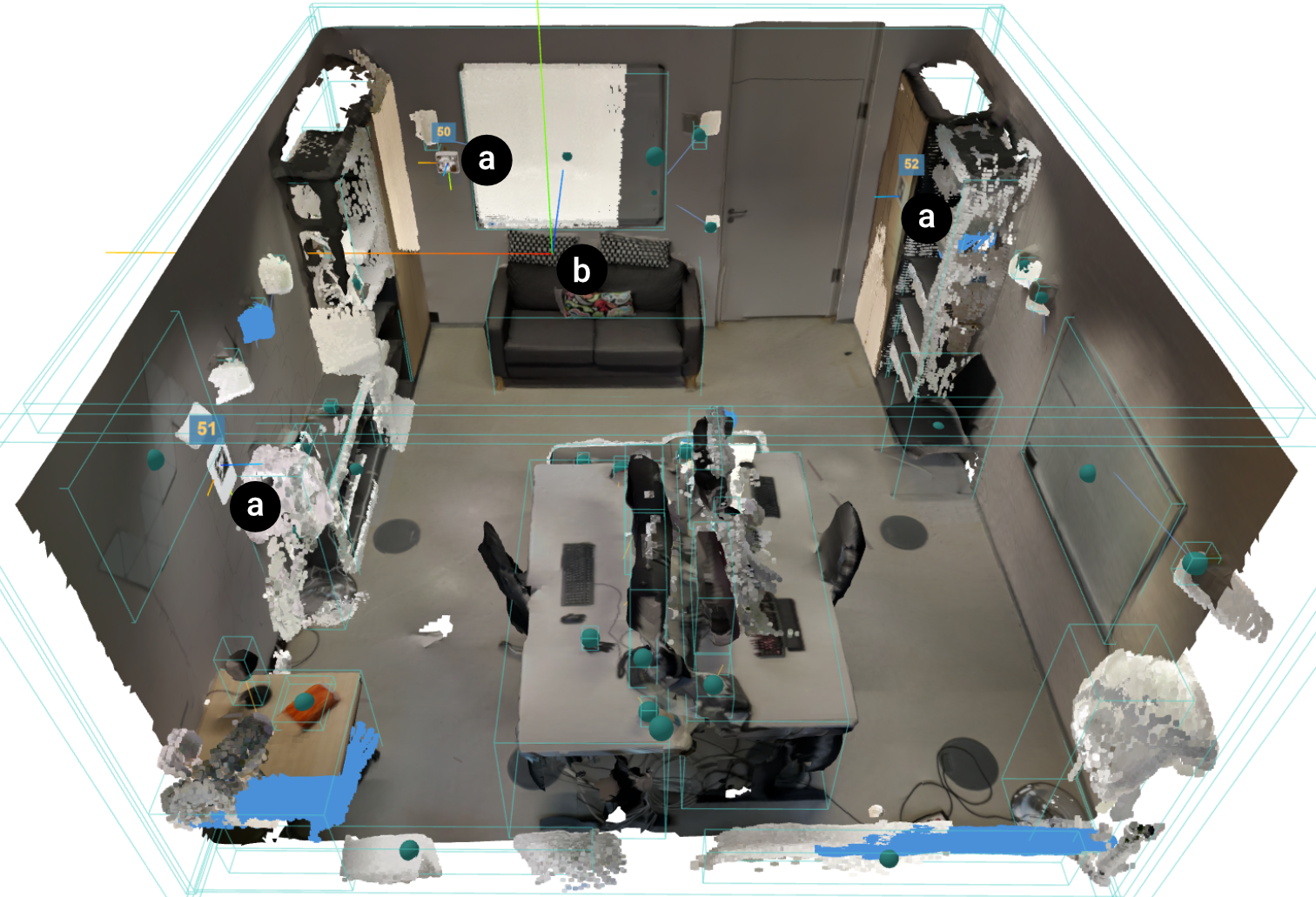}
  \caption{SSG alignment using multiple ArUco markers (a): Averaging all detected markers into one central transform (b) reduces the effect of individual marker-tracking inaccuracies.}
  \label{fig:ssg_alignment}
\end{figure}

\paragraph*{Scene overview} returns a compact summary of the largest objects in the room and their key spatial relationships (e.g., ``monitor on top of desk''). This cheap first call is usually sufficient for the agent to plan placement of virtual content.

\paragraph*{Graph queries} let the agent search the objects (nodes) and edges (relationships) of the SSG for specific tags (e.g., all objects tagged ``chair'') or relationships (e.g., ``on top of'', ``next to''), returning numeric node identifiers, bounding boxes, and semantic labels.

\paragraph*{Node lookup} returns the full record for a single object, including its geometry and all relationships in which it participates. The agent uses this tool once it has identified a target object and wants to confirm its geometry and relations. At runtime, the numeric node id can be used to fetch the live object coordinates.

\paragraph*{IoT device lookup} maps a device name (e.g., ``smart lamp'') to its API description and the numeric SSG node id of the physical object to which it is bound.

\paragraph*{Example retrieval} matches the user's request against a curated catalog of the sample applications and templates, and returns the most relevant ones for the agent to study before writing any code. This tool provides best practices for implementing recurring capabilities that the agent would otherwise have to construct from scratch with the risk of an incorrect result. The example collection includes general-purpose XR capabilities (loading 3D models, spatial audio, hand-gesture interaction, plane detection), but also capabilities specific to dealing with the physical environment (loading an SSG, marker alignment, device control).

The generated application does not hard-code coordinates or device addresses. Since SSG queries are also available to the vibe-coded application at runtime, there is no need to hard-code object positions, network addresses of IoT devices, or other data items that could become stale. Instead, the vibe-coded application contains queries to the live SSG to obtain these data items, making it resilient against many unforeseen changes in the environment, such as furniture that has been moved.

\subsection{Verification}
\label{sec:verification}

The agent is instructed to perform mandatory verification steps while generating the code. The verification tool loads the current vibe-coded project in a headless browser and reports back any runtime signals it produces (console errors and warnings, uncaught exceptions, unhandled promise rejections, and failed module imports). In addition to these error signals, it returns the live three.js scene graph of the running application, i.e., the hierarchy of objects with their types, names, geometry, materials, and world positions. With the three.js scene graph, the agent can check if the intended 3D content is actually present in the scene, which would otherwise represent a silent error. Moreover, the agent uses the SSG to perform partial feasibility checks even before writing any code. If a requested object does not exist in the scene or no IoT device is associated with it, the agent reports the problem to the user rather than producing code that is bound to fail at runtime.

Verification is embedded in the main vibe-coding loop. The agent is required to iterate as long as verification problems are found. In the next iteration, the agent inspects the reported code portion or scene content, edits the offending elements, and re-runs verification until no further problems are reported or a maximum number of iterations is exceeded.

The same loop also guards against an application that runs without errors, yet does not do what the user asked. Before finishing, the agent derives an explicit checklist of the concrete features implied by the original request. It enumerates each requested object, interaction, spatial anchor, and device action, and confirms that each one is actually present in the three.js scene graph. For example, a requested panel must exist, something must be drawn, buttons must be wired to event handlers, and device commands must correspond to a real IoT capability. A feature that is requested but missing (or only partially implemented) is treated exactly as an error and triggers another edit-and-verify iteration.

The agent terminates after the project loads cleanly and covers the request. Runtime validation is therefore woven into the generation loop rather than deferred until after generation. With this process, the agent is able to catch problems that cannot be diagnosed from source code alone, such as missing imports, exceptions during scene setup, an empty scene graph, or a request that is only partially fulfilled.

\begin{figure}[t]
  \centering
  \includegraphics[width=\smallfigwidth]{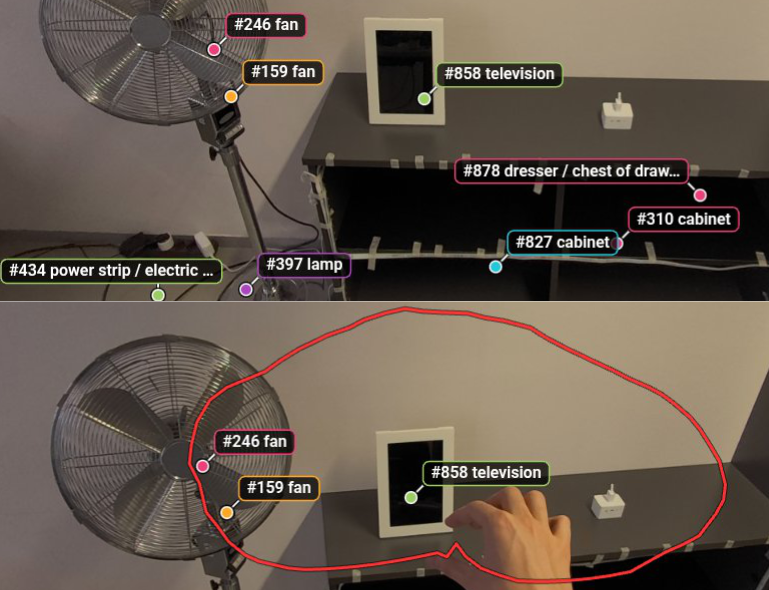}
  \caption{Set-of-mark based overlay of SSG node ids and tags on a user-triggered screen capture, used to ground follow-up refinement requests in the user's current view. Right shows the option to encircle a particular region of interest.}
  \label{fig:set_of_mark}
\end{figure}

\begin{figure*}[t]
  \centering
  \includegraphics[width=\linewidth]{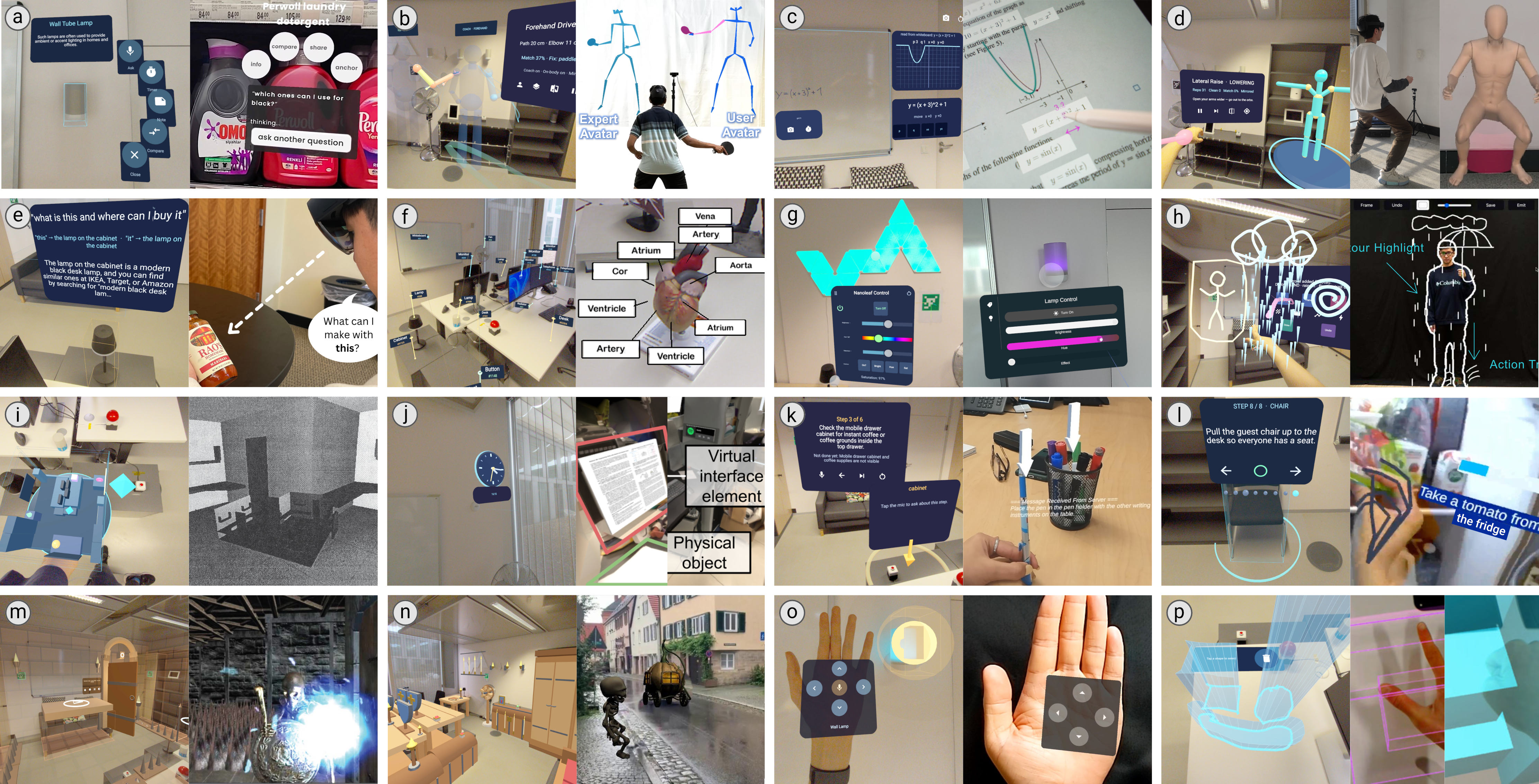}
  \caption{Collage of recreated XR papers. The left part is always our recreation, and the right is sourced from the original paper. (a) XR-Objects~\cite{dogan2024xrobjects}, (b) avaTTAR~\cite{ma2024avattar}, (c) Augmented Math~\cite{chulpongsatorn2023augmentedmath}, (d) AR-Enhanced Workouts~\cite{wu2023arworkouts}, (e) GazePointAR~\cite{lee2024gazepointar}, (f) Hedgehog Labeling~\cite{tatzgern2014hedgehog}, (g) LORIOT~\cite{kolberg2026loriot}, (h) RealityCanvas~\cite{xia2023realitycanvas}, (i) World in Miniature~\cite{stoakley1995wim}, (j) SemanticAdapt~\cite{cheng2021semanticadapt}, (k) XaiR~\cite{srinidhi2024xair}, (l) PaperToPlace~\cite{chen2023papertoplace}, (m) VRoamer~\cite{cheng2019vroamer}, (n) TransforMR~\cite{kari2021transformr}, (o) AdhocUI~\cite{du2022adhocui}, and (p) MRTouch~\cite{xiao2018mrtouch}.}
  \label{fig:paper_recreation_collage}
\end{figure*}

\subsection{Object grounding}
\label{sec:spatial_grounding}

During code generation, the user's follow-up requests must be tied to the part of the scene to which they refer. Commonly, the user observes a concrete spatial situation, but a plain text prompt does not convey the spatial aspects to the vibe-coding agent. Therefore, we enable users to point to the part of the scene they want to change. We let the user wearing the XR headset perform a ``capture'' gesture to record a snapshot of the video pass-through in the XR headset (Figure~\ref{fig:set_of_mark}). The gesture can select the entire view or a part of it by drawing an outline with the tracked hand. The snapshot is attached to the user's refinement request and passed to the vibe-coding agent. The agents is able to understand spatial references from the user request to the snapshot using set-of-mark prompting~\cite{yang2023setofmark}. We overlay the id and semantic tag of every scene object visible in the snapshot. Consequently, the snapshot communicates not only what the user was looking at, but also which nodes are in the user's view. In addition to the set-of-mark snapshot, we append a JSON representation of the visible nodes and their spatial placement to the prompt. With these preparations, spatial references in a follow-up request can be resolved into known object identifiers without having to describe spatial arrangements in words.

\section{Technical evaluation}
\label{sec:tech_eval}

In order to evaluate the capabilities of our agent, we replicate popular XR papers and perform a manual evaluation of 120 test prompts.

\subsection{Paper Recreation}
\label{sec:paper_recreation}
Inspired by Zhu et al.~\cite{zhu2025agentar}, who re-implemented six published AR applications with their agent, we demonstrate the breadth of our approach by recreating the core contribution of the six papers recreated by Zhu et al.~\cite{zhu2025agentar} and additional 10 state-of-the-art XR papers selected from classic and recent CHI, UIST, IEEE VR, ISMAR, and TVCG contributions. Each system was recreated with the agent driven by our best-performing LLM (Claude Opus 5); the full natural-language prompts are given in the supplementary material. Figure~\ref{fig:paper_recreation_collage} shows the recreations. For each paper, we summarize the core concept and state what our recreation is capable of.

\begin{itemize}[leftmargin=*, itemsep=2pt, parsep=0pt, topsep=2pt]
  \item[a] \textbf{XR-Objects}~\cite{dogan2024xrobjects} turns uninstrumented real objects into interactive ones by anchoring a contextual action menu onto each detected object. Our recreation resolves a tapped object through the SSG rather than segmentation in the camera image; the anchored menu answers LLM questions about the object and pins a running timer to it.
  \item[b] \textbf{avaTTAR}~\cite{ma2024avattar} trains table-tennis strokes with on-body and detached visual cues. Our recreation places a semi-transparent coach avatar performing the target stroke beside the user, overlays the target arm-and-paddle motion onto the tracked hand, and mirrors the user's own motion on a second avatar that highlights deviations.
  \item[c] \textbf{Augmented Math}~\cite{chulpongsatorn2023augmentedmath} turns static textbook formulas into explorable AR explanations. Our recreation allows users to write mathematical equations onto a whiteboard and spawn plots and control sliders next to it that visualize and explore the influence of certain parameters.
  \item[d] \textbf{AR-Enhanced Workouts}~\cite{wu2023arworkouts} proposes AR visual cues over ``workout at home'' videos. Our recreation shows a looping virtual trainer with a target-pose skeleton and motion-direction cues, compares the user's tracked movement against the example, and highlights deviations. The trainer iterates through various yoga poses and animates their correct movement.
  \item[e] \textbf{GazePointAR}~\cite{lee2024gazepointar} resolves ambiguous pronouns in voice queries from gaze and pointing. In our recreation, one can ask questions about certain objects in the room, and the application will respond textual and verbally to the user request like ``What is this and where can I buy it from?''.
  \item[f] \textbf{Hedgehog Labeling}~\cite{tatzgern2014hedgehog} computes overlap-free external labels along 3D poles in object space. Our recreation attaches a leader-line label to every SSG object, moving along a 3D pole, and keeps the placement overlap-free and temporally coherent while the user walks.
  \item[g] \textbf{LORIOT}~\cite{kolberg2026loriot} binds IoT devices to SSG nodes to make the IoT localization-ready. Our recreation highlights every device in the room with a sphere; clicking it opens a device-capability-fitting control panel generated from the device's capabilities that actuates the real hardware.
  \item[h] \textbf{RealityCanvas}~\cite{xia2023realitycanvas} augments real-world motion with responsive scribble animations. Our recreation moves the sketching from the touchscreen into mid-air 3D (a feature listed as future work by the original authors) and supports the main functions proposed in the original paper (i.e., object binding, motion trails, particle emitters, touch-triggered bursts, flip-book frames, and contour highlights).
  \item[i] \textbf{World in Miniature}~\cite{stoakley1995wim} is the classic handheld miniature proxy of the surrounding world. Our recreation builds a view-anchored miniature of the room from the SSG. Manipulating an object in the miniature, which is virtual in our case, moves its full‑scale counterpart, and the interaction is bidirectional.
  \item[j] \textbf{SemanticAdapt}~\cite{cheng2021semanticadapt} adapts AR widgets' layouts along virtual--physical semantic connections. Our recreation places each widget at the semantically fitting SSG location (e.g., the wall where the clock would hang, sticky notes on the desk, media controls next to the speaker) and adapts the layout when the scene or task changes.
  \item[k] \textbf{XaiR}~\cite{srinidhi2024xair} couples a multimodal LLM with XR headsets for step-by-step task guidance anchored in the world. In our recreation, the LLM generates the instruction steps, each step's panel and arrow anchor to the required object through the SSG, and the live camera view verifies completion before advancing.
  \item[l] \textbf{PaperToPlace}~\cite{chen2023papertoplace} spatializes instruction documents by placing each step at the relevant real location. Our recreation anchors the steps of a recipe to the matching SSG objects and shows one step at a time. Our instruction topic was centered around a meeting-setup checklist.
  \item[m] \textbf{VRoamer}~\cite{cheng2019vroamer} procedurally generates walkable VR experiences that fit the physical building. Our recreation generates a dungeon matching the room's free floor space, blocks the footprint of each piece of furniture with a themed hazard (e.g., spike traps, a skeleton knight), and places the dungeon door at the real doorway.
  \item[n] \textbf{TransforMR}~\cite{kari2021transformr} composes alternate mixed realities by substituting real objects with pose-matched virtual counterparts. Our recreation adapts the street-scale concept to the room. SSG objects are replaced by size-matched stand-ins fitted to their bounding boxes, so the chair becomes a throne, the desk a workbench, and the lamp a torch. Their real-time inpainting module was not reproduced.
  \item[o] \textbf{Ad hoc UI}~\cite{du2022adhocui} lets users summon opportunistic interfaces onto everyday objects and remap their functionality with voice commands and tapping gestures. Our recreation replicates the remappable controller pad from the paper's teaser. Four arrow buttons on the open palm are pressed with the other hand. Voice commands reassign the pad to another object, for example, to control the real lamp's brightness or the photo frame's images, with the currently mapped device highlighted through the SSG.
  \item[p] \textbf{MRTouch}~\cite{xiao2018mrtouch} adds multitouch input to AR by turning bare environment surfaces into touchscreens. Our recreation identifies the touch planes from the SSG rather than from fused depth and infrared imagery. The figure illustrates one of the proposed examples, in which drawn shapes are extruded into 3D space.
\end{itemize}

\subsection{Evaluation of code quality}

To assess the quality of our coding agent, we tested it on the 60 prompts of the VCXR60 dataset from Du et al.~\cite{du2026vibe}. Furthermore, we added 60 new prompts that are related to our physical test environment (VCSSG60). These prompts were deliberately phrased using object and device categories (e.g., “a lamp”, “the desk”, “all fans in the room”) rather than specific instances. Thus, they can be reused in compatible environments. We validated 480 XR projects generated by the agent using four LLM variants, each with a medium reasoning effort, from the 120 (= 60 + 60) evaluation prompts. For each test case, we recorded the generation time, the prompt- and completion-side token usage, the number and types of tool calls, the console errors and warnings thrown at runtime, and the number of verification passes the agent required before finishing. We also measured frame rate, frame times, draw calls, and submitted triangles per frame.
Frame stability indicates whether an XR application runs stably on a standalone device.
For the first 60 prompts from the VCXR60~\cite{du2026vibe} dataset, all SSG tools are disabled because these prompts are not grounded in a real environment. For the remaining 60 prompts, the SSG and IoT tools were fully enabled.

\begin{figure}[t]
  \centering
  \includegraphics[width=\smallfigwidth,height=0.72\textheight,keepaspectratio]{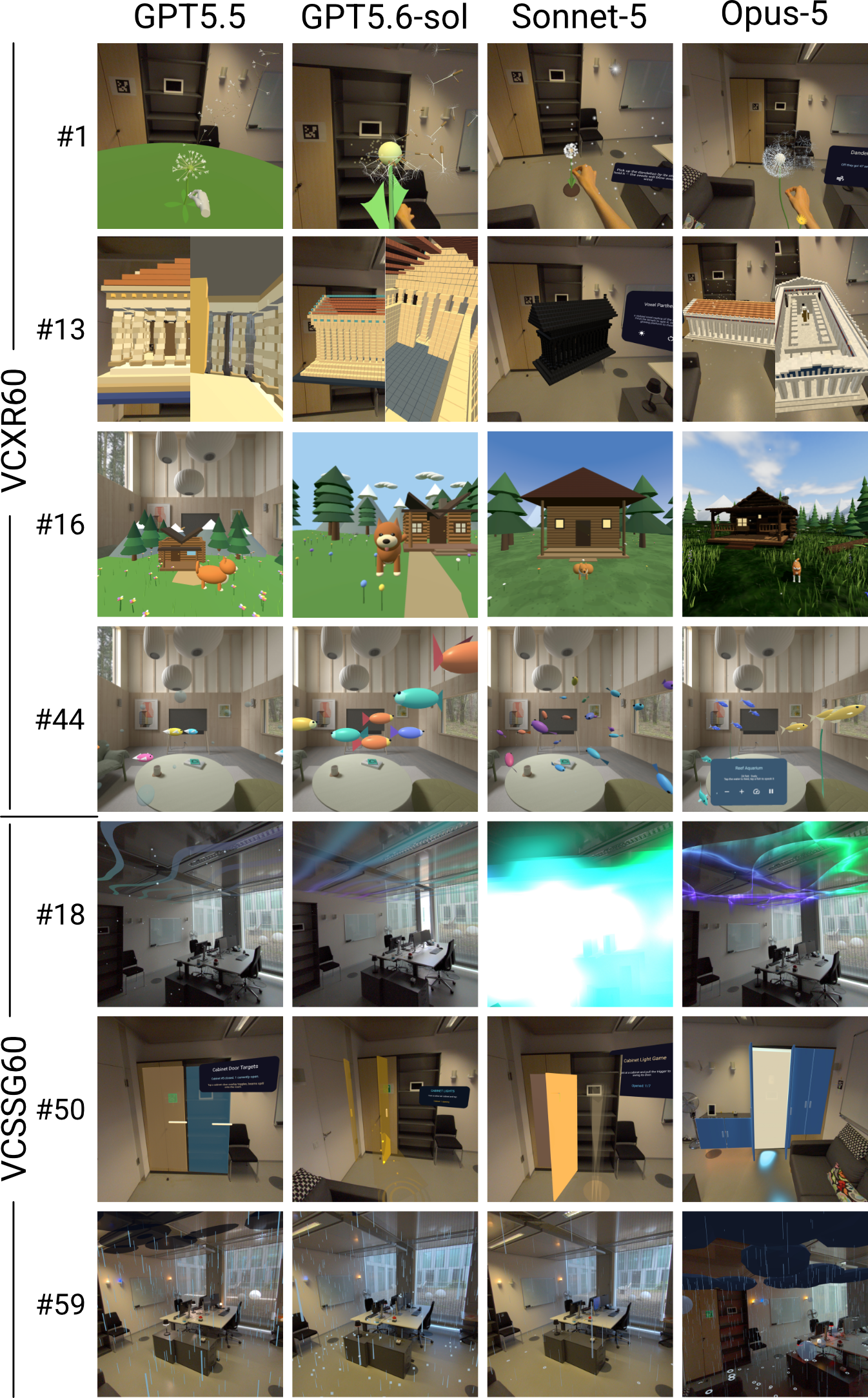}
  \caption{Seven of the 120 evaluation prompts, each vibe-coded by all four evaluated models. Claude Opus 5 is clearly ahead in visual quality and in the polish of unrequested features: It places a Zeus statue inside its Parthenon (VCXR60 \#13) and renders the Shiba Inu in front of the alpine cabin with dense grass and scene lighting (VCXR60 \#16). VCSSG60 \#50 exposes a limit of the environment model: The SSG does not specify which cabinets have doors, which face carries them, or how large they are, so every model has to guess; Claude Sonnet 5 misjudged the height of the cabinet doors.}
  \label{fig:evaluation_samples}
\end{figure}

\begin{table*}[t]
  \centering
  \resizebox{\textwidth}{!}{
\begin{tabular}{@{}llccccccccccc@{}}
  \toprule
  \textbf{Model} & \textbf{Set} & $n$ & \textbf{C1} & \textbf{C2} & \textbf{C3} & \textbf{C4} & \textbf{C5} & \textbf{C6} & \textbf{C7} & \textbf{C8} & \textbf{Overall} & \textbf{$\geq$4} \\
  \midrule
  GPT-5.5 & VCXR60 & 60 & 4.8 (0.5) & 4.7 (0.7) & 4.5 (1.0) & 4.7 (0.6) & 4.8 (0.6) & 4.3 (1.2) & -- & 4.0 (0.7) & 4.6 (0.8) & 89\% \\
   & VCSSG60 & 60 & 4.9 (0.5) & 4.6 (0.7) & 4.3 (1.2) & 4.1 (0.9) & 4.4 (0.7) & 4.7 (0.6) & 4.3 (1.4) & 4.1 (0.8) & 4.5 (0.8) & 89\% \\
  \addlinespace
  GPT-5.6 sol & VCXR60 & 60 & 4.9 (0.3) & 4.2 (0.8) & 4.6 (0.9) & 4.8 (0.5) & 4.9 (0.5) & 4.1 (1.4) & -- & 4.0 (0.8) & 4.6 (0.8) & 90\% \\
   & VCSSG60 & 60 & 4.9 (0.3) & 4.5 (0.8) & 4.6 (0.8) & 4.7 (0.6) & \textbf{4.4 (0.6)} & 4.7 (0.6) & 4.7 (0.7) & 4.4 (0.7) & 4.6 (0.7) & 92\% \\
  \addlinespace
  Claude Sonnet 5 & VCXR60 & 60 & 4.9 (0.5) & 4.0 (1.0) & 4.4 (1.2) & 4.8 (0.6) & \textbf{5.0 (0.0)} & \textbf{4.4 (1.2)} & -- & 3.9 (0.8) & 4.5 (0.9) & 86\% \\
   & VCSSG60 & 60 & 4.9 (0.3) & 4.3 (0.9) & 4.5 (0.9) & 4.4 (0.8) & 4.2 (0.8) & 4.7 (0.6) & 4.6 (0.9) & 4.0 (0.7) & 4.4 (0.8) & 88\% \\
  \addlinespace
  Claude Opus 5 & VCXR60 & 60 & \textbf{5.0 (0.3)} & \textbf{5.0 (0.2)} & \textbf{4.9 (0.3)} & \textbf{5.0 (0.2)} & \textbf{5.0 (0.0)} & 4.4 (1.0) & -- & \textbf{4.8 (0.5)} & \textbf{4.9 (0.3)} & \textbf{98\%} \\
   & VCSSG60 & 60 & \textbf{5.0 (0.2)} & \textbf{4.9 (0.3)} & \textbf{4.9 (0.4)} & \textbf{4.9 (0.3)} & 4.2 (0.8) & \textbf{5.0 (0.2)} & \textbf{4.8 (0.6)} & \textbf{4.9 (0.4)} & \textbf{4.8 (0.5)} & \textbf{97\%} \\
  \bottomrule
\end{tabular}
}
  \vspace{4pt}
  \caption{Manual quality ratings, reports at \textit{mean} (\textit{SD}) on a 5-point scale, per model and prompt set for the criteria C1--C8. Not-applicable criteria are excluded from the means; -- marks cells without any applicable rating. \textit{n} is the number of evaluated projects and $\geq$4 is the share of all ratings in the top two categories. If several models are rated on the same set, the best value per column is printed in bold.}
  \label{tab:tech_eval_quality}
\end{table*}

\begin{table}[t]
  \centering
  \resizebox{\linewidth}{!}{
\begin{tabular}{@{}lcccccccc@{}}
  \toprule
  \textbf{Model} & $n$ & \textbf{1st clean} & \textbf{Final clean} & \textbf{Time (min)} & \textbf{Tool calls} & \textbf{Verifications} & \textbf{Tok.\ in (k)} & \textbf{Tok.\ out (k)} \\
  \midrule
  GPT-5.5 & 120 & 94\% & 98\% & 3.0 (0.9) & 27.7 (8.7) & 2.2 (1.3) & 669.2 (305.2) & 8.1 (1.7) \\
  GPT-5.6 sol & 120 & 93\% & 98\% & 3.1 (0.8) & 28.2 (9.0) & 2.0 (1.0) & 543.0 (228.3) & 7.0 (1.4) \\
  Claude Sonnet 5 & 120 & 88\% & 94\% & 8.1 (1.7) & 20.6 (12.1) & 1.9 (0.8) & 1022.1 (596.9) & 9.0 (4.3) \\
  Claude Opus 5 & 120 & 94\% & 94\% & 8.9 (1.4) & 33.6 (13.0) & 2.2 (0.9) & 1796.1 (941.1) & 12.1 (7.1) \\
  \bottomrule
\end{tabular}
}
  \vspace{4pt}
  \caption{Automatic generation metrics per model over all \textit{n} vibe-coded projects: share of projects for which first and final headless verification ran without console errors, generation time, tool calls, verification passes, and token usage per project, given as \textit{mean (SD)}. Input tokens comprise the full prompt side, including cache reads and writes. Providers cache differently (Anthropic's explicit caching leaves fresh input near zero), so non-cached input alone is not comparable across models.}
  \label{tab:tech_eval_auto}
\end{table}

\begin{table}[t]
  \centering
  \resizebox{\linewidth}{!}{
\begin{tabular}{@{}llcccccc@{}}
  \toprule
  \textbf{Model} & \textbf{Mode} & $n$ & \textbf{FPS} & \textbf{Frame ms (P50)} & \textbf{Frame ms (P95)} & \textbf{Draw calls} & \textbf{Tri.\ (k)} \\
  \midrule
  GPT-5.5 & Desktop & 120 & 47.0 (12.3) & 21.8 (8.8) & 27.9 (11.1) & 75.4 (127.4) & 213.8 (176.9) \\
  \addlinespace
  GPT-5.6 sol & Desktop & 120 & 51.6 (10.3) & 19.0 (5.8) & 24.8 (9.0) & 73.9 (126.3) & 191.1 (134.8) \\
  \addlinespace
  Claude Sonnet 5 & Desktop & 120 & 53.0 (9.7) & 18.4 (6.2) & 24.9 (10.5) & 48.2 (52.6) & 181.6 (116.3) \\
  \addlinespace
  Claude Opus 5 & Desktop & 120 & 51.6 (10.2) & 18.6 (6.4) & 25.0 (9.9) & 62.7 (55.6) & 194.8 (107.6) \\
  \bottomrule
\end{tabular}
}
  \vspace{4pt}
  \caption{Frame-timing metrics of the running vibe-coded projects per model and session mode: frames per second, median and 95th-percentile frame times, draw calls, and triangles per frame. One data point per session (median over its samples), aggregated as \textit{mean} (\textit{SD}).}
  \label{tab:tech_eval_perf}
\end{table}

Due to the lack of a standardized evaluation scheme for LLM-generated applications and automations, we followed an approach similar to those used in prior work~\cite{warburton2025IoTWhispererFramework}.
Following the criteria-based assessment of LLM-generated Node-RED automations by Warburton et al.~\cite{warburton2025IoTWhispererFramework}, we rate every vibe-coded project against a fixed set of criteria rather than holistically. We designed the rating scheme to separately assess concerns that are often conflated in holistic judgments. A project may, for example, score high on visual quality but low on spatial grounding, or high on prompt adherence but low on runtime robustness. We therefore evaluated each project according to the following criteria:
%
\newcommand{\crit}[1]{\item[#1]}
\begin{itemize}[align=left, labelwidth=2.2em, labelsep=0.5em, leftmargin=2.7em,
    topsep=2pt, partopsep=0pt, itemsep=1pt, parsep=0pt]
  \crit{C1} \textbf{Prompt adherence}: All explicit and implied requirements of the prompt are present and correct.
  \crit{C2} \textbf{Visual quality}: The generated content forms a cohesive, polished look.
  \crit{C3} \textbf{Interactivity}: The intended interactions reliably trigger the correct events.
  \crit{C4} \textbf{Spatial user interface and ergonomics}: Panels and controls are legible, reachable, and comfortably placed.
  \crit{C5} \textbf{Runtime robustness}: The project can be experienced without runtime issues and maintains a stable frame rate in a real XR session.
  \crit{C6} \textbf{Spatial grounding and object alignment}: Content anchors to the correct real object, and are correctly positioned and oriented.
  \crit{C7} \textbf{IoT integration}: The correct device is resolved and actuated through valid commands.
  \crit{C8} \textbf{Polish and elaboration}: Production detail beyond what is explicitly requested in the prompt, including secondary motion, feedback on every interaction, animations, sounds, and sensible start, end, and reset states.
\end{itemize}
%
The agent verifies its own output before finishing by loading each project in a headless browser and iterating until it runs without console errors. Manually assessing basic executability would therefore largely duplicate this automatic verification procedure and provide little variation across projects. We therefore used manual rating for the two aspects that cannot be assessed through a brief execution in a headless, non-interactive, non-immersive environment:
Whether its spatial interface is usable in a headset at all (C4), and whether a project can be experienced in a real XR session without runtime issues and with a stable frame rate (C5).
Every project involving an XR interaction (e.g., gesture control) was manually tested in XR using a Samsung Galaxy XR.
Projects that did not require AR functionality were tested in a desktop browser. The authoring guide provided to the agent did not include these evaluation criteria to prevent the vibe-coded projects from being overfit to our ratings.

Not every criterion was applied to every prompt. Spatial grounding is meaningful only for 80 of 120 prompts that reference real geometry, and IoT integration is only for the nine prompts that require actuating or reading of a device. All ratings were determined by one person.
Each criterion was rated through a web-based review dashboard that displays the vibe-coded project alongside its source code, and for every run, the automated metrics were shown above. Scores were persisted per session, so the same prompt evaluated with multiple models retains a separate score set for each run.

Table~\ref{tab:tech_eval_quality} reports the manual quality ratings per model, separated into the VCXR60 prompts and our 60 environment-related prompts (VCSSG60).
The correctness criteria sit close to the ceiling for every model: Prompt adherence (C1) lies between 4.8 and 5.0---the expected consequence of a verification loop that all models share---and runtime robustness (C5), between 4.2 and 5.0. Performance is more varied on the appearance criteria: On VCXR60, visual quality (C2) spans a full point (4.0 for Claude Sonnet 5 versus 5.0 for Claude Opus 5), and polish (C8) ranges from 3.9 to 4.8. Claude Opus 5 leads almost every column, with 98\% (VCXR60) and 97\% (VCSSG60) of its ratings in the top two categories. \figurename~\ref{fig:evaluation_samples} shows where this difference comes from. Claude Opus 5 invests in unrequested but fitting detail, such as the statue inside its Parthenon or the vegetation around its alpine cabin, where the other models often settle for plain primitives. The environment-grounded prompts are handled as reliably as the baseline set: Spatial grounding (C6) lies between 4.7 and 5.0, and IoT integration (C7), between 4.3 and 4.8 on VCSSG60; only interactive robustness dips slightly (C5, 4.2--4.4 versus 4.4--5.0 on VCXR60), since grounded prompts tend to demand more interaction with the room.

Table~\ref{tab:tech_eval_auto} summarizes the automatically recorded generation metrics on all vibe-coded projects, and Table~\ref{tab:tech_eval_perf} reports the frame-timing measurements of the running projects.
The quality has a price (Table~\ref{tab:tech_eval_auto}). The OpenAI models finish a project in about three minutes, and the Claude models take eight to nine. Claude Opus 5 consumes more than three times the tokens of GPT-5.6 sol (1.8M versus 0.5M per project). First-pass cleanliness ranges from 88\% to 94\% and rises to 94--98\% after self-repair, with roughly two verification passes per project for every model. The frame-timing metrics (Table~\ref{tab:tech_eval_perf}) show no systematic performance gap. All models average 47--53 FPS with 95th-percentile frame times of 25--28\, ms. The large standard deviations of the draw-call and triangle counts, however, reveal individual outlier projects that submit an order of magnitude more geometry than the average, which supports rating frame stability per project rather than per model.

\section{User study}
\label{sec:user_study}
To further assess the agent's ability to generate spatially aware and IoT-integrated XR applications, we conducted an expert user study with 10 participants. The study was approved by the ethics committee of our institution.

\subsection{Study design}
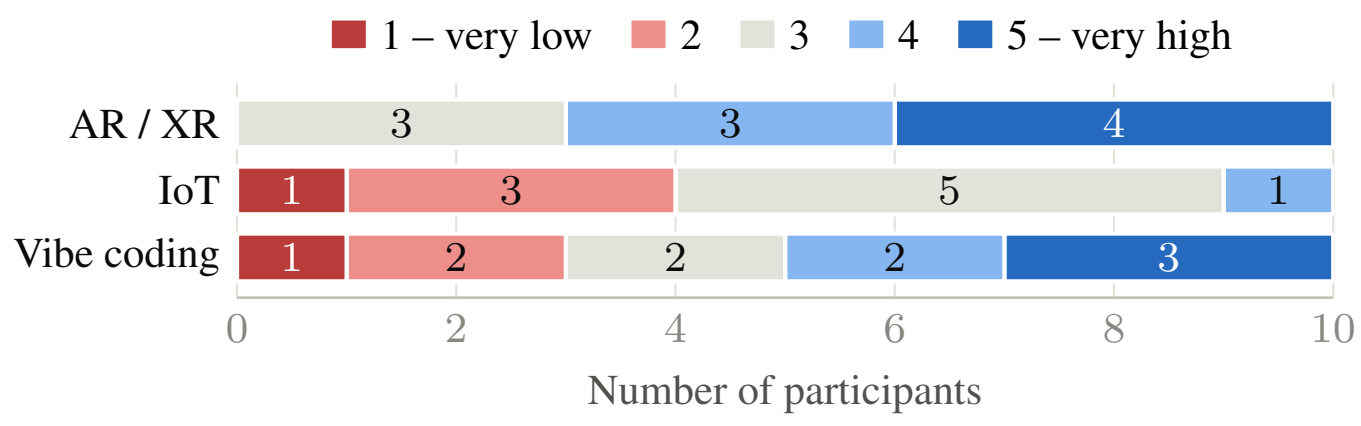
\begin{figure}[t]
  \centering
\begin{tikzpicture}
  \begin{axis}[
      likert bar chart,
      height=3.0cm,
      symbolic y coords={VC, IoT, AR},
      yticklabels={Vibe coding, IoT, AR / XR},
      ymin={[normalized]-0.6}, ymax={[normalized]2.6},
    ]
    \addplot[likert1] coordinates {(1,VC) (1,IoT) (0,AR)}; 
    \addplot[likert2] coordinates {(2,VC) (3,IoT) (0,AR)}; 
    \addplot[likert3] coordinates {(2,VC) (5,IoT) (3,AR)}; 
    \addplot[likert4] coordinates {(2,VC) (1,IoT) (3,AR)}; 
    \addplot[likert5] coordinates {(3,VC) (0,IoT) (4,AR)}; 
    \legend{1 -- very low, 2, 3, 4, 5 -- very high}
  \end{axis}
\end{tikzpicture}
  \caption{Self-reported domain experience of the 10 experts on a five-point scale (1 = very low, 5 = very high) for AR/XR, IoT, and vibe coding.}

  \label{fig:domain_experience}
\end{figure}

The study was designed to capture qualitative aspects, such as the perceived usefulness of environment-aware and IoT-integrated capabilities, and the overall user experience. We invited 10 experts (seven male, three female, 26--52 years old, $M = 32.2$, $SD = 8.2$) with knowledge in a variety of domains, covering XR, IoT and agentic coding. Figure~\ref{fig:domain_experience} summarizes their self-reported experience.

Each session lasted 90 minutes, divided into a 15-minute tutorial, a 60-minute main session, and a 15-minute closing interview. During the main session, participants interacted with the coding agent wearing a Samsung Galaxy XR headset through four openly themed prompt scenarios. We deliberately did not prescribe what to implement. Within each scenario, participants were free to choose the exact prompts and projects themselves. At the same time, the shared set of scenarios kept the experiences comparable between the participants. The scenarios were defined as follows:

\begin{itemize}
  \item Situated content on real objects: The application must attach content, such as an information display, ambient decoration, or spatial annotation, to specific real-world furniture or objects.
  \item Room-scale XR game: The application must incorporate at least two real-world objects into the gameplay.
  \item IoT control: The application must incorporate at least one IoT device.
  \item Open task: No additional requirement is imposed.
\end{itemize}
Throughout the study, the coding agent was driven by GPT-5.6 sol with a medium reasoning effort, kept identical between all participants and scenarios. Following the tutorial, each participant worked through the four scenarios in sequence, creating one project per scenario prompt. After each generation, they rated on a five-point Likert scale (1 = failure, 5 = full success) how well the agent achieved their desired outcome. Once the final project was created, participants completed the System Usability Scale (SUS) and the HALIE surveys, and the session concluded with a recorded verbal interview (transcribed with Whisper) to collect qualitative feedback.

\subsection{Quantitative results}
\label{sec:user_study_results}

During the 40 generation sessions (10 participants $\times$ 4 scenarios), participants rated 24 test cases as full success (5 on the 5-point scale), and, 11 test cases with a 4. In other words, 87.5\% of all tests landed in the top two categories ($M = 4.45$ overall). Broken down by scenario, the mean ratings were 4.3 for situated content, 4.5 for the room-scale game, 4.6 for IoT control, and 4.4 for the open task. The single worst rating (2) was given by P4 for a pointing-based IoT control application in which the ``bounding boxes were visible, but pointing was not usable''. The deficiencies were typically resolved within one or two follow-up prompts. P1 noted that, for the situated-content task, ``the alignment between the aquarium and the white board was not perfect, but a second prompt [\ldots] fixed it'', and P3 found it ``really great that one could fix problems by adding an additional specification''.

We logged every generation session, including per-state timings, prompt counts, and token usage. On average, a project took 12.8 minutes ($SD = 5.4$, range 4.0--23.7) from the first prompt to the accepted result, with very similar means across the four scenarios (situated content: 12.3, game: 13.4, IoT control: 13.3, open task: 12.3 minutes). The projects required a median of two prompts (maximum of five), and 28 of the 40 projects were completed with at most two prompts. The 11 projects accepted after a single prompt took only 6.8 minutes on average, so the total task time was dominated by the number of refinement iterations, rather than by the initial generation.

The state breakdown shows where this time went: Participants spent 30.7\% of the task time experiencing the running application in XR; 9.6\% reviewing the agent's responses, and, 44.1\% waiting for the agent to finish generating. P7 suggested a more collaborative workflow that involves the user during generation. In terms of resources, a project consumed on average 75k input tokens and 10k output tokens.

\begin{figure}[t]
  \centering
\begin{tikzpicture}
  \begin{axis}[
      likert bar chart,
      height=3.6cm,
      symbolic y coords={Ease, Satisfaction, Enjoyment, Helpfulness},
      ymin={[normalized]-0.6}, ymax={[normalized]3.6},
    ]
    \addplot[likert1] coordinates {(0,Ease) (0,Satisfaction) (0,Enjoyment) (0,Helpfulness)}; 
    \addplot[likert2] coordinates {(0,Ease) (1,Satisfaction) (0,Enjoyment) (0,Helpfulness)}; 
    \addplot[likert3] coordinates {(0,Ease) (0,Satisfaction) (1,Enjoyment) (0,Helpfulness)}; 
    \addplot[likert4] coordinates {(1,Ease) (5,Satisfaction) (2,Enjoyment) (4,Helpfulness)}; 
    \addplot[likert5] coordinates {(9,Ease) (4,Satisfaction) (7,Enjoyment) (6,Helpfulness)}; 
    \legend{1 -- very low, 2, 3, 4, 5 -- very high}
  \end{axis}
\end{tikzpicture}
  \caption{HALIE ratings of the 10 participants on a five-point scale (1 = very low, 5 = very high) for perceived helpfulness, enjoyment, satisfaction, and ease of use.}
  \label{fig:halie_results}
\end{figure}
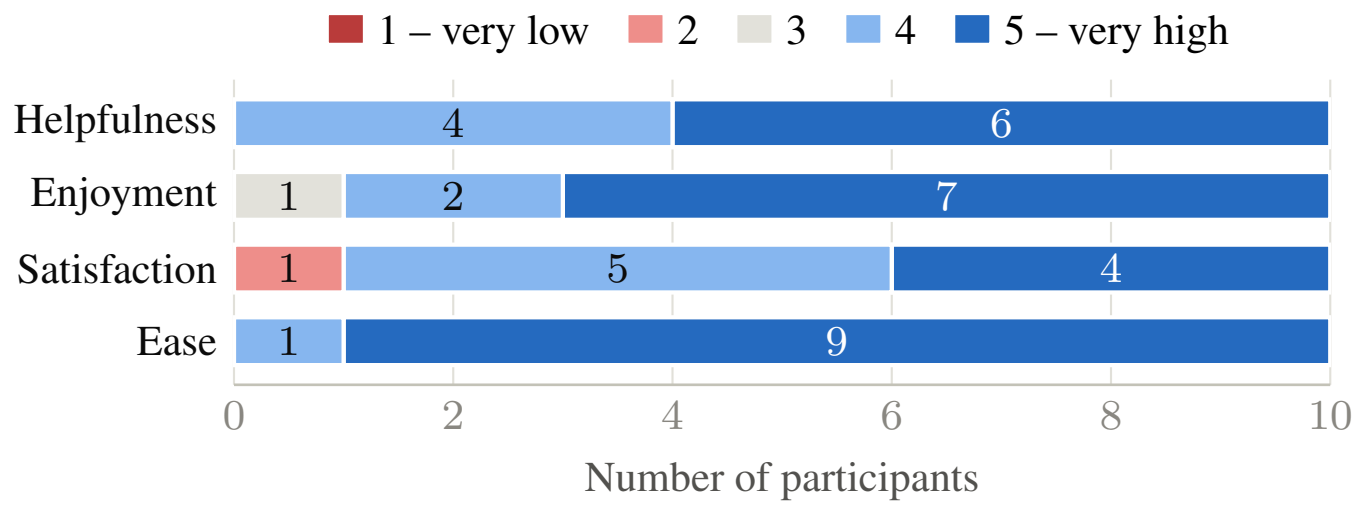

The system reached a mean SUS score of 85.25 ($SD = 9.31$, range 67.5--95). The raw NASA TLX subscales (1 = very low, 21 = very high) indicate a low overall workload: Mental demand ($M = 4.7$), physical demand ($M = 2.2$), temporal demand ($M = 2.0$), effort ($M = 3.0$) and frustration ($M = 3.5$) all stayed in the lower fifth of the scale, while perceived performance was rated high ($M = 17.7$ of 21). The HALIE ratings (Figure~\ref{fig:halie_results}) confirm this impression: Ease of use received the highest scores ($M = 4.9$, $SD = 0.3$), followed by helpfulness ($M = 4.6$, $SD = 0.5$), enjoyment ($M = 4.6$, $SD = 0.7$) and satisfaction ($M = 4.2$, $SD = 0.9$). The only satisfaction rating below 4 came from P4, consistent with the interaction problems reported above.

\subsection{Qualitative results}
We transcribed the closing interviews and grouped the statements into recurring themes.

\textbf{First attempts already close to the intent.}
Participants were consistently surprised at how well the very first generation met their request. P4 stated that ``each first iteration already captured like 80, 90\% of what I wanted'', and P6 reported that a requested \textit{onsen} atmosphere with integrated devices ``basically worked pretty much first try''. P3 found the results ``impressively good'' considering ``how little I specified how those things should work'', and P7 was ``really impressed'' that the agent implemented physics ``more or less correctly''. P10, initially skeptical that ``maybe the output would be really bad'', reported that after the first task they ``expected even more from the system afterward''.

\textbf{Spatial grounding and IoT integration were the standout capabilities.}
The environment awareness introduced by the SSG was repeatedly emphasized. P1 was ``positively surprised how well the static interior of the room was integrated, like the whiteboard [\ldots] dimensions'', and P5 highlighted ``especially the placement on the whiteboard and that you can really interact with the room''. P7 noted that ``the interaction with the IoT devices and the spatial connection worked really well'', and P4 confirmed that ``placing the bounding boxes at the [\ldots] correct real world object locations [\ldots] worked flawlessly'' and that ``controlling the IoT devices worked''. An unexpected agent behavior was observed by P6: When asked to display beach images on the physical picture frames, the agent ``overlayed an XR item on top of'' the frames instead of using the real IoT displays.

\textbf{Trust grew through iteration.}
Most of the participants described a calibration process rather than blind trust. P1 ``expected that not everything worked perfectly'' on the first try but that ``after two iterations [\ldots] things would improve and work out''. P6 (``now that I've used it, I would trust the system'') and P9 (``after the first experiment, I already saw it was capable'') converged on trust after early successes, while P8 leaned on prior vibe coding experience: ``I trust them now [\ldots] I know usually stuff works''. P4 argued that up-front verification is impractical in XR anyway (``for any complex scene, debugging [\ldots] is basically impossible without trying it out''), and P7 remained the most reserved, noting that ``testing is not easy'' and ``for a real application some more [\ldots] systematic testing would be required''.

\textbf{Faster than traditional XR development, at the cost of control.}
Compared to engine-based workflows, participants unanimously perceived a large speed-up. P2 stated ``I don't even think about using Unity now. This is definitely more efficient'', P5 found it ``way faster, more convenient'', and P10 estimated ``for 90\% of the stuff, I am faster''. The flip side was reduced control and code ownership. P7 remarked that ``in Unity you have control of everything [\ldots] with a vibe-coded application like this, there's no chance to understand the code'', and P4 raised the question of how to ``pick up from there'' when transitioning a vibe-coded project into long-term manual development. P10 similarly suspected that small visual adjustments ``would be faster in Unity, because I could just do it''.

\textbf{Envisioned use: rapid prototyping.}
When asked about future use, most participants positioned the system as a prototyping and ideation tool. P1 could ``imagine it as some kind of sandbox [\ldots] for rapid prototyping'', P5 saw value ``for prototyping [\ldots] because it's saving a lot of time and it's already mapped to the room'', and P6 called it ``pretty good for a first draft''. P3 additionally envisioned end-user smart-home configuration, i.e., ``configuring diverse programs in your home or in your office''. P9 concluded that it ``makes the whole process much easier and saves a lot of time''. P8 tied adoption to asset quality: ``the output needs to look better'', suggesting the agent should ``look for free and open source 3D models [\ldots] instead of creating trees out of primitives''.

\textbf{Limitations and requested improvements.}
The most frequent criticism concerned the visual fidelity of the primitive-based output (P1, P3, P4, P5, P8), followed by the robustness of fine-grained interactions: P4 could not reliably pick up and throw a virtual basketball or point at devices, P6 initially struggled with the pinch gesture, and P10 frequently mis-clicked the voice-recording button. Participants tended to attribute these control problems at least in part to themselves rather than to the system. P4 speculated ``maybe my prompts were not good enough'', P6 reported that the pinch issues disappeared with practice (``I got used to it [\ldots] towards the end, I got along with it much better''), and P10 framed the mis-clicks as something that ``was a bit hard for me'' rather than a fault of the system. P10 also exposed a limit on the granularity of the SSG, as the bounding-box representation does not model the free space beneath a table, which broke a hide-and-seek game. The per-session marker scanning was perceived as friction, with P1 and P6 asking for a persistent spatial configuration.
Beyond fixes, participants proposed a more collaborative agent that asks clarifying questions rather than relying on one-shot generation (P3, P7). They suggested that such an agent should handle queries about missing data, such as location or calendar access (P8),
support interactive debugging of the running scene through the agent (P4),
transparency about system capabilities before generation (P2), direct in-scene manipulation of generated objects (P4, P10), and access to the live camera feed for context-aware applications (P8).

\textbf{Emergent creativity.}
Several participants highlighted moments where the agent went beyond the literal prompt. P1 asked for a ``funny-looking skeleton'' and the agent ``actually put sunglasses on the skeleton'', and later, given a single ``Don't panic!'' reference, it ``added some kind of hidden references to those books all around the spacecraft''. P2's pink-elephant application was rated ``super cute and working perfectly'', and P6 was surprised ``how well the primitives can replicate the actual objects'', including animated steam bubbles.

\section{Discussion}
\label{sec:discussion}

Taken together, the technical evaluation and the expert study indicate that grounding a vibe coding agent through an SSG moves XR application creation from a manual, editor-bound activity to a conversational one. The experts produced working, spatially anchored, IoT-connected applications in a median of two prompts and consistently positioned the system as a rapid prototyping tool. At the same time, both evaluations expose where the current prototype ends and what separates it from production use. We discuss these themes in the following.

\textbf{Asset fidelity bounds the use cases.}
The visual quality of the primitive-based output was the most frequent criticism in the user study (P1, P3, P4, P5, P8) and is the main reason the participants framed the system as a prototyping rather than an authoring tool. This ceiling is not inherent to the approach: Retrieving open 3D assets instead of composing primitives (P8), together with textures, lighting, and shadows, would directly raise it, and the same holds for audio (P1 wished for sound effects) and for networked capabilities such as multiplayer or cloud services. The spatial user interface is bound the same way. The framework offers a small set of panel, grid, and button components, and every interface the agent produces is assembled from them. A richer user interface library would directly widen what a vibe-coded interface can look like. Since the agent already bases its code on retrieved examples, extending asset retrieval is a natural next step.

\textbf{From one-shot generation to a collaborative loop.}
The current interaction model is essentially one-shot. The user prompts, waits, and inspects the result. Instead, participants asked for an agent that poses clarifying questions before committing to one interpretation (P3, P7), is transparent about its capabilities before generating (P2), and can be queried about the running scene for debugging (P4). The task-time analysis reinforces this direction: Participants spent 44\% of the task time waiting for the agent (Section~\ref{sec:user_study_results}), and the verification loop that drives the high final code correctness is precisely what makes code generation slow. A collaborative workflow would not only reduce misinterpretations but would also mask this latency, e.g., through streamed intermediate results or background planning while the user is still exploring the scene (P4). Masking the wait may matter as much as shortening it.

\textbf{The environment model needs more detail, persistence, and an explicit grounding policy.}
Several observed failures are traced back to the environment model rather than to the agent, such as the coarse bounding boxes and the need for fiducial marker scanning. The same holds for the semantic depth of a node. The SSG records what an object is and where it stands, but not which face of a cabinet carries its doors, in which direction that face points, or what kind of doors they are. The agent must guess such details, and guesses differ per model (Figure~\ref{fig:evaluation_samples}, VCSSG60 \#50). Enriching nodes with oriented part-level attributes would remove this guesswork. Alternatively, a more detailed SSG could include not just a cabinet but also its doors and their hinge orientation. IoT command mapping proved to be reliable, but device discovery and status feedback are the obvious next steps. More fundamentally, the agent currently lacks an explicit policy for when to activate a physical device (e.g., it used a virtual overlay instead of a real screen). Making this choice explicit or allowing the agent to ask would prevent such silent substitutions.

\textbf{Verification, not model choice, drives output quality.}
In the technical evaluation, 92\% of the projects loaded cleanly on the agent's first verification pass, and 96\%, after self-repair, confirming the execute--observe--fix loop as the main quality driver. Across the four LLM variants evaluated, first-pass cleanliness (88--94\%), generation time (3--9 minutes) and token usage (0.5M--1.8M prompt-side tokens per project) differ substantially, but all models converge to comparably runnable projects through the same loop. The only areas where they differ are visual quality and polish (Table~\ref{tab:tech_eval_quality}). The loop, however, only observes load-time behavior in a headless browser. Runtime performance, such as frame rate and draw calls, ergonomics, and robustness while the project is actually experienced, remains invisible to it. This is exactly why our rating assesses these aspects manually (C3--C5). Closing this gap, e.g., by feeding the frame-timing probe back into the agent, would extend self-repair from ``it runs'' to ``it runs well''.

\textbf{Code ownership and scale remain open.}
Finally, experts questioned how a vibe-coded project transitions into long-term development. P7 noted that ``there's no chance to understand the code'' of a vibe-coded application, P4 asked how to ``pick up from there'' when continuing manually, and P10 doubted that the approach scales beyond small projects. Two directions appear viable: keeping the user permanently on the agent level, so that generated code is never touched by hand, or exporting vibe-coded projects into an editor-friendly structure that tools and humans can maintain. Which of the two best supports applications that grow over months is an open question for future work.

\textbf{Missing standardized evaluation schema.}
How to evaluate vibe-coded XR applications is not yet settled. agentAR~\cite{zhu2025agentar} rated generation quality with an LLM as judge: GPT-4 assessed whether the agent's \emph{plan} satisfied the user's request, complemented by the precision and recall of the selected tools against ground-truth plans. Such proxy measures never observe the resulting application itself, and agentAR's generation loop closes only over compiler errors, so runtime behavior remains unmeasured; its authors themselves name a workflow in which the agent can inspect and test the generated application as future work. Closer to our position, Warburton et al.~\cite{warburton2025IoTWhispererFramework} score LLM-generated IoT automations against fixed per-criterion checklists, which our rating scheme adopts and extends to the XR-specific concerns of Section~\ref{sec:tech_eval}. We evaluate by executing instead: every vibe-coded application is run during generation (Section~\ref{sec:verification}), which directly observes runtime errors and the resulting scene, and the aspects that a brief automated execution cannot show are rated manually in the headset against fixed criteria (Section~\ref{sec:tech_eval}). We consider direct execution combined with such criteria a step toward a standardized evaluation schema for vibe-coded XR applications.

\section{Conclusion}

We presented a vibe coding agent that generates spatially aware and IoT-integrated WebXR applications by grounding generation in an SSG of the user's actual environment. The scene graph, enriched with IoT device bindings, serves the agent as a queryable tool at authoring time. The application generated with vibe coding uses the SSG as a live data source at runtime. A verification loop ensures that the projects run before they reach the user. Across the recreation of 16 published XR systems, a technical evaluation on 120 prompts, and an expert study with 10 participants, the approach produced working, spatially anchored applications in minutes and a median of two prompts, and was consistently perceived as a fast alternative to engine-based prototyping. Future work will follow the directions identified in our discussion: higher-fidelity assets and a richer spatial user interface library, a collaborative generation loop that hides latency behind interaction, a finer and persistent environment model, and feeding runtime performance back into the agent's self-repair.

\section*{Acknowledgments}

This work was supported by the Alexander von Humboldt Foundation, funded by the German Federal Ministry of Research, Technology and Space, the German Research Foundation \textit{DFG} under Germany's Excellence Strategy EXC 2120/2 (390831618), and the Austrian Science Funds \textit{FWF} (I5912).

\bibliographystyle{abbrv-doi}

\bibliography{references}

\end{document}